\documentstyle[12pt,aasms4]{article}
\begin{document}
\setcounter{secnumdepth}{4}
\def\arcsec{\ifmmode^{\prime\prime}\;\else$^{\prime\prime}\;$\fi}
\def\arcmin{\ifmmode^{\prime}\;\else$^{\prime}\;$\fi}
\def\kms{~km~s$^{-1}$}
\title{The Progenitor Masses
of Wolf-Rayet Stars and
Luminous Blue Variables
Determined from Cluster Turn-offs. I. Results from 19 OB Associations in
the Magellanic Clouds}

\author{PHILIP MASSEY\altaffilmark{1}}
\affil{Kitt Peak National Observatory, National Optical Astronomy
Observatories\altaffilmark{2}
\\ P.O. Box 26732, Tucson, AZ 85726-6732}

\author{ELIZABETH WATERHOUSE\altaffilmark{3} and
KATHLEEN DEGIOIA-EASTWOOD}
\affil{Department of Physics and Astronomy, Northern Arizona University,
P.O. Box 6010, Flagstaff, AZ 86011-6010.}

\altaffiltext{1}{Visiting Astronomer, Cerro Tololo Inter-American
Observatory, National Optical Astronomy Observatories, which is operated by
the Association of Universities for Research in Astronomy, Inc.\ (AURA) under
cooperative agreement with the National Science Foundation.}
\altaffiltext{2}{Operated by AURA
under cooperative agreement with the
National Science Foundation.} 
\altaffiltext{3}{Participant in Research Experiences
for Undergraduates program, Northern Arizona University, Summer 1998. 
Present address:
Harvard University, 436 Eliot House Mail Center, 101 Dunster St., Cambridge, MA 02138.}

\begin{abstract}

We combine new CCD {\it UBV} photometry and spectroscopy with that from the
literature to investigate 19 Magellanic Cloud OB associations that contain
Wolf-Rayet (WR) and other types of evolved massive stars.  
Our spectroscopy reveals a wealth of newly identified
interesting objects,
including early O-type supergiants, a high mass double-lined binary in the
SMC, and, in the LMC, a newly confirmed LBV (R~85), a newly discovered
WR star (Sk$-69^\circ$194), and a newly found luminous B[e] star (LH85-10).  
We use these data to provide
precise reddening determinations and construct physical H-R diagrams for
the associations. We find that about half of the associations may be
highly coeval, with the massive stars having formed
over a short period ($\Delta \tau <$ 1~Myr).  The (initial) masses
of the highest mass {\it unevolved} stars in the
coeval clusters may be used to estimate the masses of the progenitors of
WR and other evolved stars found in these clusters.
Similarly the bolometric luminosities of the highest mass unevolved stars
can be used to determine the bolometric corrections for the evolved
stars, providing a valuable observational basis for evaluating recent 
models of these complicated atmospheres.
What we find is the following: 
(1) Although their numbers are small, it appears that the WRs
in the SMC come from only the highest mass ($>70 \cal M_\odot$) stars.
This is in
accord with our expectations that at low metallicities only the most massive
and luminous stars
will have sufficient mass-loss to become
WRs.  
(2) In the LMC, the early-type WN stars (WNEs) occur in clusters
clusters whose turn-off masses range from 30$\cal M_\odot$
to 100 $\cal M_\odot$ or more.  
This suggests that possibly all stars with mass $>30 \cal M_\odot$ pass
through an WNE stage at LMC metallicities.
(3) The one WC star in the SMC is found in a cluster
with a turn-off mass of 70$\cal M_\odot$,  the same as for the SMC WNs.
In the LMC, the WCs
are found in clusters
with turn-off masses of 45$\cal M_\odot$ or higher, similar to what is found
for the LMC WNs.  Thus we conclude that WC stars 
come from essentially the same mass range
as do the WNs, and indeed are often found in the same clusters.  This
has important implications for interpreting the relationship between
metallicity and the WC/WN ratio found in Local Group galaxies, which
we discuss. (3) The LBVs in our
sample come from very high mass stars ($>85 \cal M_\odot$), similar to what is known
for the Galactic LBV $\eta$~Car, suggesting that only the most massive stars
go through an LBV phase.  Recently, Ofpe/WN9 stars have been implicated
as LBVs after one such star underwent an LBV-like outburst.  However,
our study includes two Ofpe/WN9 stars,
BE~381 and Br~18,  which we find in clusters with much lower turn-off masses
($25-35 \cal M_\odot$). We 
suggest that Ofpe/WN9 stars 
are unrelated to ``true" LBVs: not all ``LBV-like outbursts" may have the
same cause. Similarly, the B[e] stars have sometimes been
described as LBV-like.  Yet, the two stars in our sample appear to come
from a large mass range ($>30-60 \cal M_\odot$). This is
consistent with other studies
suggesting that B[e] stars cover a large range in bolometric luminosities.
(4)
The bolometric corrections of early WN and WC stars are found to be 
extreme, with an average BC(WNE)=$-6.0$~mag, and an average BC(WC4)=$-5.5$~mag.
These values are considerably more negative than those of even the hottest
O-type stars.  However, similar values have been found for WNE stars by
applying Hillier's
``standard model" for WR atmospheres.  We find more modest BCs for the
Ofpe/WN9 stars (BC=$-2$ to $-4$~mag), also consistent with recent analysis done
with
the standard model. Extension of these studies
to the Galactic clusters will provide insight into how massive stars evolve
at different metallicities.

\end{abstract}


\section{Introduction}

Conti (1976) first proposed that Wolf-Rayet (WR) stars might be a normal, late
stage in the evolution of massive stars.
In the modern version of the 
``Conti scenario" (Maeder \&
Conti 1994), strong stellar
winds gradually 
strip off the H-rich outer layers of the most massive stars during the course
of their
main-sequence lifetimes.  
At first the H-burning CNO products He and N are revealed,
and the star is called a WN-type WR star; this stage occurs either near the
end of core-H burning or after core-He burning has begun, depending upon
the luminosity of the star and the initial metallicity. Further mass-loss
during the He-burning phases exposes the triple-$\alpha$ products C and O, and results in a WC-type WR star. 
Since the fraction of mass that a star loses during
its main-sequence evolution depends upon luminosity (mass), we
would expect that at somewhat lower masses evolution proceeds only
as far as the WN stage.  At still lower masses a star never loses 
sufficient mass to become a Wolf-Rayet at all, but spends its
He-burning life as a red supergiant (RSG). Mass-loss rates also scale with
metallicity as the stellar winds are driven by radiation 
pressure acting through highly ionized metal lines. Thus the 
mass-limits for becoming WN or
WC stars should vary from galaxy to galaxy, and with location within a galaxy
that has metallicity variations.  

Studies of mixed-age populations in the galaxies of the Local Group have 
confirmed some of the predictions of the Conti scenario. For instance, the
number ratio of WC and WN stars is a strong function of metallicity
(Massey \& Johnson 1998 and references therein), with proportionally more WC stars seen at higher
metallicities, suggesting that the mass-limit for becoming WC stars is somewhat
lower in these galaxies.  Similarly the relative number of WRs and RSGs is  correlated with metallicity, and there
is a paucity of high luminosity RSGs at high metallicities (Massey 1998a),
suggesting that these high luminosity stars have become WRs rather than RSGs.

However, fundamental questions remain concerning the evolution of massive stars:

\noindent
(1) What is the role of the luminous blue variables (LBVs)?  These stars are
highly luminous objects that undergo photometric ``outbursts" associated with
increased mass-loss (Humphreys \& Davidison 1994).  Are LBVs a short
but important stage in the lives of {\it all} high mass stars that occur at
or near the end of core-H burning? Recent efforts have linked some of
the LBVs to binaries, as Kenyon \& Gallagher (1985) first suggested. 
The archetype LBV, $\eta$ Car,  may be a binary with a highly eccentric orbit
(Damineli, Conti, \& Lopes 1997), but whether its outbursts
have anything to do with the binary nature remains controversial
(Davidson 1997), as does the orbit itself (Davidson et al.\ 2000).
Similarly, the WR star HD~5980 in the SMC underwent an
``LBV-like" outburst (Barba et al 1995); this star is also
believed to be a binary with an eccentric orbit, although the nature
(and multiplicity?)
of the companion(s) remains unclear (Koenigsberger et al.\ 1998; Moffat 1999).

\noindent
The
Ofpe/WN9 type WRs, and the high-luminosity B[e] stars have recently
been implicated in the LBV phenomenon. 
The former have spectral
properties intermediate between ``Of" and ``WN" (Bohannan \& Walborn 1989).
One of the prototypes of this class, R~127, underwent an LBV outburst in 1982
(Walborn 1982; Stahl et al.\ 1983; see discussion in Bohannan 1997).  
Similarly some B[e] stars have been described
as having LBV-like outbursts.  Var~C, a well-known LBV in M~33,
has a spectrum indistinguishable from B[e] stars: compare
Fig.~8a of
Massey et al.\ (1996) with Fig.~8 of Zickgraf et al. (1986).   
Do all B[e] stars undergo an LBV phase or not? Conti (1997) has
provided an insightful review.

\noindent
(2) What is the evolutionary connection between WN and WC stars? 
We expect only the highest mass stars become WCs, while
stars of a wider range in mass become WNs.  The changing proportion of
WCs and WNs within the galaxies of the Local Group have been
attributed to the expected dependence of these mass ranges on
metallicity.  However, the relative
time spent in the WN and WC stages may also change with metallicity,
complicating the interpretation of such 
global measures drawn from mixed-age populations.

\noindent
(3) Is there any evolutionary significance to the excitation subtypes?
Both WN and WC stars are subdivided into numerical classes, or more
coarsely into ``early" (WNE, WCE) or ``late" (WNL, WCL) based upon
whether higher or lower excitation ions dominate. 
Recent modeling by Crowther (2000) suggests that the distinction between
WNL and WNE is not actually due to temperature differences
but primarily metal abundance.
Armandroff \& Massey (1991) and Massey \& Johnson (1998) have argued that
this true for the WC excitation classes based upon the metallicity
of the regions where these stars are found.

If we knew the progenitor masses of LBVs and the various kinds of WRs
we would have our answers to the above.
However, here recourse to stellar evolution models
fails us.  
Stellar evolutionary models show that a star's path in the HRD
during core-He burning is strongly dependent upon the amount of mass-loss that
has preceded this stage. Thus the nature of the LBV phenomenon becomes very
important in understanding where WRs come from, 
as the amount of mass ejected by LBVs is large, but
given the episodic nature of LBVs, hard to
include in the evolutionary models.
In addition, the locations of WRs and LBVs
in the H-R diagram are highly uncertain. LBVs have pronounced UV-excesses
and ``pseudo-photospheres" (Humphreys \& Davidson 1994).
For WR stars,  neither
the effective temperatures nor
bolometric corrections are established, as none of the standard
assumptions of stellar atmospheres hold in the non-LTE,
rapidly expanding, ``clumpy" stellar winds where 
both the stellar continua and emission-lines arise (e.g., Conti 1988).
While the WR subtypes represent some sort
of excitation sequence in the stellar winds, the relationship, if any, to
the effective temperature of the star remains unclear. 
  
There has been recent success in modeling WR atmospheres, with
convincing matches to the observed
line profiles and stellar continua from the UV to the near-IR. 
These models have the potential for determining the
bolometric luminosities and effective temperatures.
The ``standard WR model" (Hillier 1987, 1990) assumes a spherical geometry and
homogeneity, and then iteratively solves the equations for statistical
equilibrium and radiative equilibrium for an adopted velocity law, mass-loss
rate, and chemical composition. (See also Hillier \& Miller 1998, 1999.) Comparison with observations then permits 
tweaking of the parameters.  Although the solutions may not be unique, good
agreement is often achieved with observations, and in a series of papers,
Crowther and collaborators have offered the ``fundamental" parameters 
(effective temperatures, luminosities, chemical abundances, 
mass-loss rates, etc.) of WN
stars obtained with this model (Crowther, Hillier \& Smith 1995a, 1995b;  Crowther,
Smith, \& Hillier 1995c; Crowther et al. 1995d; Crowther, Smith, \& Willis 1995e; 
Crowther \& Smith 1997; Bohannan \& Crowther 1999). 

Here we utilize a complementary, observational
approach to the problem, one that can not only
answer the question of the progenitor masses of LBVs and WRs, but also provide
data on the BCs that can help constrain and evaluate the WR
atmosphere models. 

\subsection{The Use of Cluster Turn-offs}

A time-honored method of understanding the nature of evolved stars is to determine the turn-off luminosities in clusters containing such objects 
(Johnson \& Sandage 1955; Schwarzschild 1958).  
This was first applied by Sandage (1953)
to determine the masses of RR~Lyrae stars in the globular clusters M~3 
and M~92, with a result that was at variance with
that given by 
theory (Sandage 1956). 
Similarly, the turn-off masses of intermediate-age open clusters were used
by Anthony-Twarog (1982) to determine the progenitor masses of white dwarfs.
However, it is one thing to apply this to clusters with ages
of $10^{10}$ yr,
as was done for the RR~Lyrae stars, or to clusters whose ages are
$2\times 10^{7}$---$7\times 10^{8}$ yr,
as was done for white dwarfs.
Can we safely extend this 
to clusters whose ages are only of order 3--5$\times 10^6$~yr in order to
determine the progenitor masses of WRs and LBVs?

When stars form in a cluster or association, stars of intermediate mass appear
to form over a significant time span---perhaps over several million years
(Hillenbrand et al.\ 1993; Massey \& Hunter 1998). However,
modern spectroscopic and photometric studies have shown that the massive stars
tend to form in a highly coeval fashion.  For instance, in their study of the
stellar content of NGC~6611, Hillenbrand et al.\ (1993) 
found a {\it maximum} age spread of 1~Myr for the massive stars, and
noted that the data were consistent with {\it no} discernible age spread.
for all one could tell ``the highest-mass stars could have all been born
on a particular Tuesday."  Similarly, the high mass stars in the R136 cluster
have clearly formed over $\Delta \tau < 1$~Myr, given the large number of
O3~V stars and the short duration that stars would have in this phase
(Massey \& Hunter 1998).

Such short time scales for star formation are consistent with recent 
studies by Elmegreen (1997, 2000a, 2000b), who argues 
that star formation takes place not over tens of crossing times but 
over one or two.  
For regions with large spatial extent (such as 100~pc
diameter OB associations) star formation in the general region may occur
over a prolonged time ($\leq$10~Myr).  
However, large OB associations can contain subgroups that have formed independently
(Blaauw 1964), and are small enough so that a high degree of coevality
($<1-2$~Myr) is expected.  The stars from such a subgroup need not be
spatially coincident.  Rather, a star with a random motion of 10~km~s$^{-1}$
will have traveled 30~pc in just 3~Myr. Thus in an OB association we may
find intermediate-mass stars which have formed from a number of
subgroups over time, but  massive stars which may have
formed from a single subgroup and hence are coeval---even though these 
massive stars may now be spread out throughout the OB association.
Or, it may be that massive stars of different
ages are present, in which case the ``turn-off mass" will not be relevant
to the evolved object.  We take an optimistic approach in our search for
turn-off masses, but will insist that coevality be established empirically
for the massive stars in the region in question.

For massive stars, the mass-luminosity relationship is much flatter than for
solar-type stars ($L\sim M^{2.4}$ for 30~$\cal M_\odot$ and $L\sim M^{1.5}$ for
120~$\cal M_\odot$). As a result, the lifetimes of massive stars do not change as drastically with mass as one might expect.  A 120 $\cal M_\odot$ will have a main-sequence
lifetime of 2.6~Myr, a 60 $\cal M_\odot$ still will have a main-sequence lifetime of 3.5~Myr, and a 25 $\cal M_\odot$ star will have a main-sequence
lifetime of 6.4~Myr. (These numbers are based on the $z=0.02$ models of Schaller et al.\ 1992.)

Thus it should be possible to use clusters and OB associations to pin down the
``minimum mass" of various unevolved massive stars.  If the highest mass
star still on the main-sequence is 60$\cal M_\odot$, and its associated
stellar aggregate contains a WC-type WR star, then we might reasonably
conclude that the progenitor mass of the WC star was at least 60 $\cal M_\odot$.
Of course, if coevality does not hold, then this answer may be wrong---the WC star might have come from a 25 $\cal M_\odot$ that formed earlier.  But were that the case, 
it would have to have formed {\it much} 
earlier---at least 3~Myr earlier, according to the lifetimes given above,
and such an age spread should be readily apparent.

We can in principle also find the BCs from the cluster turn-offs.  
It is straightforward to 
determine the absolute visual magnitude of the WR, making
some modest correction for the emission lines.
Since massive stars evolve at nearly constant bolometric luminosity,  
we expect that the bolometric luminosity of the WR
will be at least as great as the bolometric luminosity of the highest
mass main-sequence object.  With modern stellar models we can improve on this
by making first-order correction for modest luminosity evolution.

We are, of course, not the first to have trod on this ground.  
Schild \& Maeder
(1984) attempted to provide links between the different WR 
subtypes using this sort of analysis of Galactic clusters, concluding that
stars with masses as low as $18 \cal M_\odot$ became WN stars, while WC stars
came from stars of $35 \cal M_\odot$ and higher, and proposing various evolutionary relationships between the various subtypes.
Humphreys, Nichols,
\& Massey (1985) also used data drawn from the literature on (mostly the same)
Galactic clusters, and found a considerably higher minimum mass for becoming
a WR star (30 $\cal M_\odot$), with no difference between the masses required
to become a WN or a WC.  They were also
the first to apply this method to determining the minimum
bolometric corrections for WR stars, concluding that WNE stars have BCs $<-5.5$~mag,
WNL stars have BCs $<-3.5$~mag, and WCs have BCs $<-5.0$~mag.  (These BCs
are considerably more negative than had been commonly assumed.)  Smith, Meynet, \& Mermilliod (1994) re-addressed the
issue of BCs by analyzing the same data from the literature on what was also
mostly the same clusters, finding BCs for WNs that were typically $-4$ mag
(WNL) to $-6$ mag (WNE), and $-4.5$ for WCs, essentially unchanged from the
Humphreys et al.\ findings.

There were
problems, however, with these earlier studies.
The most significant one was the reliance upon (the same) literature data
for the spectral types of the main-sequence stars in these clusters and
associations.
Over the past decade we have examined the stellar content of numerous clusters
and OB associations in the Milky Way, and invariably discovered stars of
high mass that had been previously missed either due to reddening or simple
oversight (Massey, Johnson, \& DeGioia-Eastwood 1995a).
A related problem is that some of the literature spectral types were
``outdated" for the O-type stars, particularly for stars of type O7 and earlier,
which would lead to an incorrect assignment of bolometric
corrections and hence luminosities and masses.
In addition, our understanding of massive star evolution has improved to
the point where we can do a considerably better job in assigning masses,
and in particular understand the errors associated with this procedure
(see, for example, Massey 1998b). Another problem was that the spectral information was sufficiently sparse that
no test of coevality could be applied to the cluster.  In addition, poor photometry---often photographic---led to poor reddening corrections.
And, finally, a significant limitation in these earlier studies
was that all were restricted to the Milky Way. 
It would be most interesting to understand the origin of evolved
massive stars as a function of metallicity; for this, extension to the
Magellanic Clouds is a logical step.

We have attempted to rectify these problems by carrying out a modern analysis
of OB associations containing WR and other evolved massive stars in galaxies
of the Local Group, obtaining new spectroscopic and photometric data where
warranted, and combining this with studies drawn form the recent literature.
In this first paper we will determine the progenitor masses of WR and LBVs
in 19 associations of the Magellanic Clouds. 
These two galaxies have 
abundances which are low compared to the solar neighborhood.
In the next paper we will compare these to
new results obtained for OB associations in our
own Galaxy. In a third paper we will combine {\it HST} photometric and
spectroscopic data with large-aperture ground-based studies to
extend this work to the more
distant members of the Local Group as an addition check on metallicity effects.

Throughout this paper we will assume the
true distance modulus of the SMC is 18.9, and that of LMC is 18.5
(Westerlund 1997; van den Bergh 2000). 

\section{Sample Selection and Observing Strategy}

In selecting this sample, we first compared the locations of known WRs and
LBVs to that of the cataloged OB associations in the SMC and LMC.
The probability of a chance supposition of a rare evolved object
against one of these associations is, of course, low.

There are nine known WR stars in the SMC (Azzopardi \& Breysacher 1979;
Morgan, Vassiliadis, \& Dopita 1991).  Four of these are within three of the
OB associations identified by
Hodge (1985). We list these in Table~1. The WR star
HD~5980 underwent an ``LBV-like outburst" in 1994 (Barba et al.\ 1995). 
This star is located in NGC~346, which is included in
our study. Three
other SMC stars described as LBV-like in some way are R~40, which is not a
member of any association: R~4, a B[e] star with ``brightness variations
typical for LBVs" (Zickgraf et al.\ 1996), located in Hodge~12, but not included
here, and AV~154 (aka S~18), another B[e] star tied to LBVs
(Morris et al.\ 1996), located just outside of Hodge~35, also not included here.
One other high luminosity B[e] star, R~50 (aka S~65=Sk~193), is listed by Zickgraf et al.\ (1986),
but is well outside any OB association.

For the LMC, Breysacher (1981) cataloged 100 Wolf-Rayet stars; an occasional
additional one has been found spectroscopically (e.g., Conti \& Garmany 1983;
Testor, Schild, \& Lortet 1993), plus components of R~136 and other
crowded clusters have been successfully resolved, which brought 
the total of known WR stars in the LMC to 134 (Breysacher,
Azzopardi, \& Testor 1999).
As part of the present study, we discovered a new WR star, Sk$-69^\circ$~194,
located in LH~81.
We compared the positions of WRs
against the Lucke-Hodge OB associations 
(Lucke \& Hodge 1970; Lucke 1972), using only those associations with
``A1" classifications.  Not all were included in the
current study; we list in Table~1 the 16 associations that are, along
with their WR stars.

Next we considered the LMC LBVs.  Six 
are listed by 
Bohannan (1997): S~Dor, R~71, R~127, HD~269582, R~110, and R~143.  To this
list we propose that R~85 be considered a seventh, based upon our discovery
here of spectral variability
(Section~\ref{Sec-r85}) and a recent characterization of its
photometric variability (van Genderen, Sterken, \& de Groot 1998; see also
Stahl et al.\ 1984).  Of these seven, S~Dor and R~85 are in LH~41, which is 
included here, and R~143 is in LH~100, which is not. We 
argue later that one of the LH~85 stars may also be an LBV based upon its spectral
similarity to other LBVs, but further monitoring is needed to establish
variability; we include it in Table~1 as a previously unknown,
high luminosity B[e] star. 
Three other ``LBV candidates" are
listed by Parker (1997) :  
R~99, 
S~61 (BE~153=Sk$-67^\circ 266$), and
S~119 (HD~269687=Sk$-69^\circ 175$).  Of these, only one is located near
an OB association (R~99 near LH~49), and it is not included here.
Finally, we also considered the location
of the high luminosity 
B[e] stars (Table~1 of Lamers et al.\ 1998;
see also
Zickgraf et al.\ 1986,  Zickgraf 1993, and 
in particular Fig.~10 in Gummersbach, Zickgraf, \& Wolf 1995).
Only S~134, 
is found in one of our regions (LH~104), although several 
are found in  other
OB associations; i.e., S~22 in LH~38 and R~82 in LH~35.

We have referred to all of these stellar aggregates as ``OB associations",
although the distinction
between an OB association, and a bona-fide ``cluster" young enough
to contain O-type stars, is hard to quantify.
The classical distinction, that clusters are gravitationally bound,
is hard to establish, as it requires
a census down to the low-mass components, plus detailed
radial velocity studies. Semantics aside, our
primary concern is to what degree these regions are coeval. Certainly
most of the OB associations studied as part of our efforts to determine the
IMFs are (Massey et al.\ 1995b).  
For the new ones studied here, we will establish the
degree of coevality directly from the data.

Our observing strategy had similarities to our work 
that determined the initial mass functions in the LMC 
(e.g., Massey et al.\ 1989a, 1995b).    
It is possible to infer masses of main-sequence
O- and B-type stars using their position in the physical H-R diagram
($\log T_{\rm eff}$ vs. $M_{bol}$) and comparing these with modern evolutionary models.  There may be systematic problems with the masses thus inferred,
although there is good agreement with the overlap of masses determined
directly from spectroscopic binaries up to 25$\cal M_\odot$ (Burkholder,
Massey, \& Morrell 1997), above which mass there is a scarcity of suitable
data on binaries.  Massey (1998b) discusses the errors in the inferred mass
with temperature; since the BC is a steep function of the effective temperature,
accurate knowledge of the latter is needed for this procedure to work. 
Sufficient accuracy cannot be achieved from photometry alone, but knowledge of
the spectral type of the star yields adequate information in most cases.
The sort of error bars associated with this can be found in Fig.~1(c) and 1(d)
of Massey et al.\ (1995b).  We will revisit this issue in Section~\ref{Sec-coeval}.

For this project we considered relying simply on the photographic photometry
or aperture photoelectric photometry that was available; e.g., Lucke (1972)
or Azzopardi \& Vigneau (1982), for the Large and Small Clouds respectively.
After all, for the stars with spectroscopy (and hence accurate BC 
determinations)
an error of 0.1 mag in the $B-V$ color will lead to a 0.3 mag error in $M_V$,
given $A_V=3.1\times E(B-V)$. An error of 0.3 mag in $M_V$ translates to an
error of 15\% in the derived mass (see details in Massey 1998b).
(For comparison, if we were relying upon the colors alone and were dealing with
a 0.1 mag uncertainty in $B-V$ we would have a 2 mag uncertainty in the BC,
and thus a 0.4 dex uncertainty in the log of the mass (i.e.,
a factor of 2.5 uncertainty in the mass of the star).

For determining the IMF, it is necessary to pursue spectroscopy down the main-sequence until spectral-type of early B or later, after which good
photometry provides as accurate information. 
Yet, in the case of determining
the turn-off masses in principle we need to only ascertain that we have obtained
spectra of the most massive unevolved object in the association.  In a 
strictly coeval population with uniform reddening, this will be equivalent to
knowing the spectral type of the visually brightest member.  However, given 
finite
photometric errors, slight non-coevality,
reddening which is spatially variable across a cluster, the
presence of other evolved supergiants (either members or field interlopers),
and the
need to demonstrate coevality, our initial aim was to obtain spectra for the
six or seven visually brightest stars in each of these associations. 
Still, this is far fewer than what would be needed to construct the
IMF.

Some of these associations had extensive CCD photometry and modern
spectroscopy in the literature, and for these we constructed H-R diagrams
and obtained a few additional spectra where warranted.   In other cases,
we already had existing unpublished CCD photometry (and in some cases
even spectroscopy) that had been aimed at determining the IMF; the complete
data for these associations, and the IMF analysis, will be published 
separately elsewhere.
For the most part, though,
we began with published photographic photometry, using this list to
select the appropriate (brightest and bluest) stars for spectroscopy, and
subsequently obtained new CCD {\it UBV} data in order to better correct
for reddening.  In all cases we examined the preliminary H-R diagrams and
then obtained spectra of the few remaining interesting stars, as needed.

\section{New Data}
We list in Table~1 the source of the data we used, be they new or from the
literature, or both.   For the new data, we identify the year in
which it was obtained.

For most of the associations (LMC) we began with the photographic iris photometry of Lucke (1972) or older sources, and
obtained spectra of the brightest and bluest stars during a run on the
CTIO 1.5-m telescope during  1996 Oct 27-31.   
Grating 58 was used in second
order with a CuSO$_4$ blocking filter, yielding wavelength coverage from
$\lambda 3750$ to $\lambda 5070$ with approximately 3\AA\ (2.8 pixels)
resolution.  The Loral chip was formated to
500 $\times$ 1200  (15--$\mu$m) pixels.  The slit was opened to 1.5 arcsec
(85$\mu$m)  and oriented EW, except for crowded regions,
where the slit angle was adjusted and/or the slit narrowed.  A typical
S/N of 100 per 3\AA\ spectral resolution 
element was achieved in a 5 min exposure
at $V=12$. 

On the night following this run (i.e., 1996 Nov 1)
we obtained {\it UBV} images of any OB associations without previous CCD
data, using the Tektronix 2048$\times 2048$ CCD imager on the CTIO 0.9-m. 
The field-of-view (FOV) was 13.5~arcmin by 13.5~arcmin, quite ample
for the typical 3 arcminute diameter OB associations in our sample. Exposure times were 
usually 100 sec in $U$ and 50 sec in each of $B$ and $V$.
The night was mostly
photometric, although the alert observing assistant reported seeing a
single cloud pass by part way through the night; later we will argue
that this affected the {\it U} photometry of two regions but nothing else.
Standard stars were 
observed at the beginning, middle, and end of the night, and reduced
satisfactorily (0.01 mag rms residuals in $U$, $B$, and $V$ in the fits to
the solutions). Nevertheless,
we treat the data as potentially non-photometric, comparing the derived
reddening-free index $Q=(U-B)-0.72\times (B-V)$ with that expected on the
basis of spectral type as a check, as described in Section \ref{Sec-hrd}.
As we discussed above, our photometric requirements are in any event modest,
given our extensive spectroscopy. 

About half of the OB associations in our sample had previously been imaged
with an RCA CCD on the CTIO 0.9-m in 1985 October by two of the present
authors (PM and KDE). The full details of these data are given in 
Massey et al.\ (1989a).  Although the FOV was only $2.5\times 4.0$ arcmin
in size, overlapping frames were taken when needed in order to 
include the whole
of an
OB association.   The photometric integrity of these 1985 data is very
high, as standard star observations were obtained over 10 photometric nights
and used for precise determinations of zero-points and color-terms. 

Similarly, some of the stars have
previously unpublished spectroscopy obtained as part of our program to
determine IMFs in the Clouds. Data obtained in 1989-1992 (Table~1) were
taken on the CTIO 4-m telescope with the RC spectrograph.  The details of
these data were given by Massey et al.\ (1995b); here we will simply note that
they were of comparable spectral resolution (3\AA), and covered at least the
wavelength region from Si~$\lambda 4089$ through He~II~$\lambda 4686$. The
S/N were typically 75 per 3\AA\ spectral resolution element.

After our preliminary HRDs were constructed, we had two
observing opportunities to obtain additional spectra where warranted.
On 1999 Jan 3-7 we used the CTIO 4-m for significantly higher resolution
and better S/N data. Grating KPGLD was used in second order
with a CuSO$_4$ filter resulting in a resolution of 1\AA\ (2.5 pixels) and
a wavelength coverage of 3730\AA\ to 4960\AA\ using the Loral
$1024\times 3100$ (15~$\mu$m) CCD.  The S/N obtained was typically 160 per
1\AA\ resolution element.
We obtained one final observation for this project on 1999 Oct 21
using the CTIO 1.5-m.  

\subsection{Analysis}

\subsubsection{Spectroscopy}
\label{Sec-spectra}

We classified the spectra with reference to the Walborn \& Fitzpatrick (1990)
spectral atlas of O and B stars. Based upon our internal consistency and previous experience
we expect that the spectral subtypes are
determined to an accuracy of one subclass and one luminosity class (e.g.,
supergiant vs giant), 
except for the earliest O-type stars, for which there is little or no
ambiguity in subclass. (See discussion in Massey et al.\ 1995a, 1995b.)

There is no metallicity dependence
in classifying hot stars as to spectral subclass, as the primary spectral type
(effective temperature) indicators are the relative strengths of different
ionization states of the same ion; e.g., He~I vs. He~II for the O-type stars,
and Si~IV vs. Si~III for the early B-type stars; however, it is our experience
that the luminosity indicators are metallicity dependent, even for the O-type
stars. This makes physical sense---in fact, it would be hard to see how this 
would fail to be the case---as the O-type luminosity indicators are primarily
indicators of the strength of the stellar wind (i.e., He~II emission vs. He~II
absorption).  The B-type luminosity indicators rely upon how strong certain
metal lines are relative to, say, He, and again we expect this to have a 
metallicity dependence. We therefore always checked the ``MK" 
luminosity class with that expected on the basis of the absolute magnitudes,
as described below; we note cases where we have adjusted the luminosity class
based upon the absolute magnitudes.

All told, we classified slightly over 200 stars.  We include our classification,
as well as those from the literature, in the catalog we describe in Section~\ref{Sec-catalog}.  Here we will illustrate and comment
on just a few of the more
interesting spectra. 

\paragraph{R~85.}
\label{Sec-r85}

We propose that the luminous star R~85 in LH~41 be
considered an LBV.  Based upon their characteristic of its {\it photometric}
variability, van Genderen et al.\ (1998) state that the star is ``undoubtedly
an active LBV."
We show in Fig.~\ref{fig:r85} some of the
{\it spectral} changes that have taken place in recent years; we
agree with van Genderen et al.'s characterization.
Feast, Thackeray, \& Wesselink
(1960) classify the star as ``B5~Iae", and note the presence of H$\beta$
emission, H$\gamma$ and H$\delta$ absorption, as well as 
its photometric variability.  Our 1996 spectra did not appear totally
consistent with this description, as Mg~II $\lambda 4481$ was present but
there was
little or no He~I~$\lambda 4471$; for a B5 star the latter should be somewhat
stronger.  We took a very high signal-to-noise spectrum with the CTIO 4-m in
January 1999, and were surprised by the rapid and strong changes present since
1996;
the newer spectrum shows the star to be hotter (based upon He~I to Mg~II)
with much stronger lines.
Dr.~B. Bohannan was kind enough to make available a photographic spectrogram
he obtained in 1985 on the Yale 1-m, along with a sensitometer exposure; there
is very good agreement between his exposure, and what we obtained 11 
years later.  The recent change in the spectrum of R~85 suggests that further monitoring would be of interest. The photometry 
listed in Table~2 comes from the 1 Nov 1996 
observation; e.g., $V=10.53$, $B-V=0.16$, and $U-B=-0.81$.  In the
1985 data (28 Nov) the star was slightly brighter: $V=10.44$, $B-V=0.12$,
and $U-B=-0.71$.

\paragraph{Newly Identified O3 Stars.} 
\label{Sec-o3s}

As part of this investigation we came across
a number of previously unrecognized O3 stars, stars whose effective temperatures
are at the extreme of the 
spectral sequence of luminous stars. We show examples
in Figs.~\ref{fig:o31} and ~\ref{fig:o32}.

First, let us consider the O3 supergiants (O3~If*) and giants
[O3~III(f*)].  These evolved stars are still in the 
temperature regime covered by the O3 classification,
and thus all such stars must be extremely massive.  
Walborn et al.\ (1999) classify the star LH90$\beta$-13 as O4~If+ on the
basis of an FOS spectrum obtained with {\it HST}, but
our higher signal-to-noise spectrum (with higher resolution) 
reveals N~V $\lambda \lambda 4603,19$
absorption; this, combined with the lack of He~I makes this an O3 star
(Fig.~\ref{fig:o31}).
The star ST5-31 in LH~101 was classified as O3~If* by Testor \& Niemela (1998);
our spectrum is in good agreement with that.
We consider the star W16-8 in LH~64 an O3~III(f*) owing to the relative
weakness of He~II $\lambda 4686$, despite the extremely strong N~IV $\lambda
4058$ emission and very strong N~V $\lambda 4603, 19$, usually indicative of
high luminosity; the absolute magnitude we derive in the next section is
$M_V=-5.4$, consistent with this classification, and reminding us that 
slight abundance anomalies can mask as luminosity effects in early-type stars.
A detailed atmospheric analysis
of this star is in progress in collaboration with Rolf Kudritzki. 

Among the O3 dwarfs (Fig.~\ref{fig:o32}) we include ST2-22 (in LH~90).
This star was previously recognized as an O3, but called a 
giant by Testor et al.\ (1993). The lack of emission at
He~II $\lambda 4686$, and the weakness of N~IV $\lambda$~4058, 
suggest a lower luminosity
class. We classify W28-23 in LH~81
as an O3~V((f)).  The star ST5-27 in LH~101 was called an O4~V by 
Testor \& Niemela (1998).  
The spectrum of this star is strongly contaminated by nebular
emission lines. We tentatively adopt an O3~V((f)) spectral type, but our data
are not inconsistent with the O4~V((f)) designation; we do not show the
spectrum as the nebular lines makes casual
comparisons difficult.
Another star in LH~81, W28-5, appears to be intermediate
between O3((f)) and O4~V((f)): the strength of He~I $\lambda 4471$ relative
to He~II~$\lambda 4542$ would argues that the star is a little bit later than
O3, but there is  N~V~$\lambda 4602,19$ present on our high signal-to-noise
spectra, and this
has usually been taken as characteristic of O3s.

The presence of He~I $\lambda 4471$ is easy to discern on the O3 stars in
Fig.~\ref{fig:o32} because of the extraordinarily high S/N (160 per 1\AA\ 
resolution element).  The O3 class was introduced 
by Walborn (1971) to describe
four stars in Carina which showed
no He~I $\lambda 4471$
on well-widened IIa-O emulsion spectrograms obtained
at modest resolution (2\AA).  When finer-grain plates were used at higher
resolution, He~I $\lambda 4471$was detected with equivalent widths of 120-250~m\AA\ by Kudritzki (1980) and Simon et al.\ (1984) for three of the
Carina stars.  Here we find that
He~I~$\lambda 4471$ lines have equivalent widths of
75~m\AA\ in W28-23, and 105~m\AA\ in ST2-22, significantly smaller than
that measured for the stars which first defined the class.  Yet modern
spectroscopy makes it possible to readily detect these lines.  

\paragraph{Other O-type Stars.}
\label{Sec-Os}

There are clearly other exceptions to the premise that N~V $\lambda 4603,19$
absorption is indicative of a luminous O3 star.  
In Fig.~\ref{fig:oms} we show the spectrum
of ST5-52, a star in LH~101 classified by Testor \& Niemela (1998) as
O3~V.  However, the strength of He~I suggests a considerably later O5.5 type.
It is easy to infer the basis for the Testor \& Niemela classification of
this star: our spectrum shows both NIV $\lambda 4058$
emission and N~V $\lambda 4603,19$ absorption, typically assumed to be
{\it only} characteristic of luminous O3 stars.
One possibility is that this star is a spectrum
binary, consisting of an O3~III(f*) plus a later O-type companion, which
contributes the He~I.  However, we propose instead that this is
 a ``nitrogen enhanced" star, and classify it as ON5.5V((f)).
We prefer this latter explanation because we have identified another LMC star,
not connected with the present study, whose He~I to He~II ratios are
consistent with an O5 type, but which also shows N~IV emission and
N~V absorption.  Detailed atmospheric analysis is underway for
both stars, pending {\it HST} data.

The star LH58-496 was classified as ``O3-4~V" by Garmany, Massey,
\& Parker (1994). Our high S/N spectrum (Fig.~\ref{fig:oms})
obtained with the CTIO 4-m shows a somewhat later spectral type,
O5V((f)).
In Fig.~\ref{fig:oms} 
we also show two other early-type dwarfs, an O5~V((f)) star
and an O4~V((f)) star.  

We illustrate a few newly discovered
luminous O-type supergiants in Fig.~\ref{fig:osgs}.  Examples shown here
include supergiants from O4 through O8.

\paragraph{A Reconsideration of Br~58 as a WR star, and A Newly Discovered
WR Star.}
\label{Sec-wrs}

The star Br~58 in LH~90, has long been recognized as
a WN Wolf-Rayet star.
Testor et al. (1993)
classify it as WN6-7, while earlier work has classified it as WN5-6 (Conti \& Massey 1989).  We illustrate its spectrum in Fig.~\ref{fig:wrs} from a new
high-dispersion, high S/N observation.
We note that our ground-based spectrum shows strong 
N~V $\lambda 4603, 19$ {\it absorption}; this, plus the considerable strength
of its absorption line spectrum, would tempt us to reclassify this as an
extreme O3~If* star, i.e., O3If*/WN6. (See Fig.~3 in Massey \& Hunter 1998.)
These stars are believed to be young, H-burning hot stars whose very high
luminosities result in sufficiently strong stellar winds to mimic the
strong emission characteristic of a WR.

The star Sk$-69^\circ194$=W28-10 in 
LH~81 is a newly discovered WR star, of type B0~I+WN.
The spectroscopic
discovery of another WR star in the LMC is not surprising, particularly
given the weakness of the emission in this object. (The equivalent width
of He~II~$\lambda 4686$ is $-2$\AA, compared to typical $-30$\AA\ for a
very weak-lined WN star; presumably this is due to the continuum 
being dominated by the
B0~I component.)
We question below whether all B0~I+WN are truly binaries.  

\subsubsection{Photometry}

{\it UBV} photometry is needed only 
(a) to determine accurate $M_V$ values for the stars with spectra,
and (b) to check that we obtained spectra for all of the likely ``most massive
unevolved star" candidates.  In order to accomplish (a)
we typically needed 
$V$ and {\it B-V} data for half a dozen stars or so in each association, and to
accomplish (b) we also required {\it U-B}, in order to construct a
reddening-free index. Nevertheless, with modern techniques it proved just as
easy to measure photometry for all stars on a frame, typically
 several {\it thousand} stars.
At least we could then be assured that the brightest stars were well-measured,
in the sense that their photometry was not contaminated by resolved neighbors.

We did this by fitting point-spread-functions (PSFs) using DAOPHOT
implemented under IRAF.  The 1996 CCD frames were measured by E.W., while the
1985 data were measured by P.M.  The method used is similar to that described
by Massey et al.\ (1989a) and we will give only a brief overview here.  Automatic
star-finding algorithms were used to identify stellar sources down to the
``plate-limit" (typically 4$\sigma$ above the noise). 
Aperture photometry through a small
digital aperture (with a diameter corresponding roughly to the full-width
at half-maximum of the stellar images) were then run in order to determine the
local sky values for each star (determined from the modal value in an annulus
surrounding each star) and to determine the instrumental magnitude to assign
to the PSF stars. For each
frame isolated, well-exposed stars were chosen to define the PSF.  This PSF
was then simultaneously fit to all of the stars whose brightnesses could 
possibly overlap.  In regions of nebulosity, the
sky value was also fit separately; otherwise, an average sky value was adopted
for all the stars in a given fitting exercise.  A frame in which the fitted
PSFs had been subtracted was then examined to see how well the PSF matched and
to look (by eye) for any stars that had been buried in the brightness of other
stars.  In addition, the $U$, $B$, and $V$ frames were blinked along with the
fitted xy centers to make sure there was consistency.  Missing stars were added
back into the star list and a final run was made on each of the three colors.
Aperture corrections were then determined for each frame in order to correct
the instrument zero-point (based upon the small digital aperture) to the large
apertures used to measure the standard stars.   These instrumental magnitudes
were then transformed to the standard system. In the case of the 1985 RCA CCD
data there were often overlapping frames involved in covering a region, and
the final photometry was combined to produce a single star list, with stars with
multiple entries averaged. 

One region, Lucke-Hodge~41, was common to
both data sets, and thus served as an end-to-end
independent check on the final, transformed photometry.  If we consider the twenty brightest
stars (in $V$) we find a mean difference 
(new minus old data set)
of $+0.015$ mag in $V$, $+0.011$ mag in $U-B$, and $+0.014$ in $B-V$, with
sample standard deviations of 0.06 mag, 
0.02 mag, and 0.04 mag, respectively. If 
two outliers are removed from the $V$ data, and one from the $U-B$ data, the
mean differences become +0.002 mag and $+0.001$ mag, respectively with
standard deviations of 0.03 mag and 0.04 mag.  This agreement is excellent,
and suggests that no systematic differences exist between the two data sets 
over the magnitude and color ranges of interest.

\subsection{The Catalog}
\label{Sec-catalog}
We list in Table~2 
the brightest stars in each of the 14 associations for which we have
new photometry; existing and new spectral types are also given.  We include all stars
of magnitudes $V=15$ or brighter; 
in several cases we extended this to fainter magnitudes
to include additional stars with spectral types or, in the case of NGC~602c, to include
at least 10 stars.  For two of the associations 
(LH~58 and LH~101) we reply upon cited
studies (cf., Table~1) but have a few new spectral types; we include these in Table~2.
(For three addition associations, NGC~346, LH~9, and LH~47, we reply purely on the cited
works in Table~1.)   

In listing the stars we make use of published names where available
finding charts exist,
although the celestial coordinates given in Table~2 should be of sufficient accuracy
to remove the need for finding charts. For the
LMC, we have kept with the star numbering given
in the finding charts of Lucke (1972), with additional stars given designations
of 1000+ so as to avoid confusion.  The exceptions are those associations with
modern CCD studies, where we have kept with the numbering scheme employed by
the authors.  In a few cases the associations contained stars that were
saturated on our CCDs (typically $V<10$); we include photometry of these 
stars from the literature.
We   describe
below details related to each association, making reference to
the results obtained in subsequent sections.

\subsubsection{Descriptions of Individual Associations}

{\it NGC~346:}  We rely on the CCD
photometry and spectroscopy of Massey, Parker,
\& Garmany (1989b).  The imaging data had their source in the same observing
run as the 1985 imaging used for many of the other associations studied
here.   Four of the brightest stars were also subjected to detailed
analysis by Kudritzki et al.\ (1989).  Reanalysis of these stars by
Puls et al.\ (1996) was used in the spectral type to effective temperature
calibration of Vacca, Garmany, \& Shull (1996), which we adopt
in the next section; we note here that despite the different methodology
involved, the masses determined by
Puls et al.\ for these stars are in good agreement with those we compute
in the following sections.
The visually brightest star is HD~5980, the WN3+abs Wolf-Rayet that underwent
an LBV-like outburst.  The second
 brightest star is the O7If star Sk~80.  More than a magnitude fainter visually are the very early O-type stars first
found by Walborn (1978), Walborn \& Blades (1986), and Niemela, Marraco, \& Cabanne (1986).

{\it Hodge~53:} Our photometry here is a comprehensive mosaic of several CCD frames and
extensive spectroscopy obtained with the goal of determining the IMF.  However, the 
the region is not condensed, and there are several stars of type
A-F and later, some of which are apparently foreground dwarfs or giants,
and others which are SMC supergiants. Our spectrum of AV~331 shows it to be an
SMC member of type
A2~I, based both on its radial velocity, appearance of the hydrogen lines, and
the strength of Fe~II $\lambda 4233$ (see Jaschek \& Jascheck 1990, Fig~10.2).
However, our spectrum of AV~339a shows it to be an F2 foreground star, probably
a dwarf, based both on its radial velocity and lack of
luminosity-sensitive Sr~II $\lambda 4077$.  A fainter star, h53-144,
is an A8 foreground dwarf.
We lack spectra for the other yellow
stars, and so we cannot comment further on their membership.
Our spectroscopy has also identified a double-lined spectroscopic
binary (O4~V+O6.5~V) which is among the most bolometric luminous members. 
When we construct the HRD, we will consider that each of the two components
contributes equally to the visual flux, consistent with the
appearance of our double-lined
spectrum, and the expected $M_V$s of stars of these spectral types.
The visually brightest member is the WR binary AV~332=Sk~108=R~31=AB~6 (WN3+O6.5) with
a 6.54 day orbit (Moffat 1982, 1988; Hutchings et al.\ 1984; Hutchings, Bianchi, \& Morris 1993).  Hutchings et al.\ (1984) argue convincingly that the O-type
companion dominates the visual flux by a factor of 10 to 1 (making it of
luminosity class ``I"),  and that its
location in the HRD suggests an initial mass of $70-80 \cal M_\odot$,
consistent too with its Keplerian mass.  Our analysis will yield a very
similar value. The other WR member, AV~336a=AB~7, is quite a bit fainter. 
The WR component is likely a WN3 (Moffat 1988), 
although all that is certain is that
it is earlier 
than WN7 (Conti, Massey, \& Garmany 1989).  An O-type absorption
spectrum is also seen. 
Recent work by Niemela (1999)
suggests
a 19.6 day period. 

{\it NGC~602c:} NGC~602 is located in the wing of the SMC; the
region was studied by Westerlund (1964), who identified three sub-components.
Components ``a" and ``b" are adjacent and are immersed in nebulosity known as
N90 (Henize 1956); component ``a" is also known as Lindsey~105 (Lindsey 1958).
Here we are concerned with the third component, ``c", which is an
isolated condensation with little nebular emission.  It was designated as a
separate association
both by Lindsey (1958) and Hodge (1985), and is known as ``Lindsey 107", and 
``Hodge~69". (See Plate 5 and Figure~1 in Westerlund 1964.) 
We obtained new CCD photometry of NGC~602c. Its visually brightest star
is the WR star AB~8, the only WC star known in the SMC.
It has enhanced oxygen, and was classified by Conti et al.\ (1989) as
``WO4 + abs". (Crowther, De Marco, \& Barlow 1998  instead 
call the WR component
``WO3".)   A new spectrum of the star obtained as part of the present program suggests
that the absorption spectrum is O4~V.   Moffat, Niemela, \& Marraco (1990) present an
orbit for this system with a period of 16.644 days. They propose spectral types of
WO4+O4~V, with which we concur, although Kingsburgh, Barlow, 
\& Storey (1995) suggest a somewhat later type for the O star.

{\it LH5:} Our photometry and spectroscopy are the first modern study of
this association. The visually brightest star is Sk~$-69^\circ30$,
a G-type supergiant according to Feast et al.\ (1960), 
with the next brightest
star an O9~I.  The WR star, Br4, was described as ``WN2" by Conti \& Massey 
(1989), as no N lines are visible, similar to the WN2 Galactic 
star HD~6327.  Like that star, Br~4 has a faint absolute magnitude.
We will find in subsequent sections that the star has a normal bolometric
luminosity, and that its faintness is presumably due to a very high temperature,
which shifts its light into the unobserved UV.
In constructing our HRD we find that the G5~Ia star Sk~$-69^\circ~30$ is 
coeval with the rest of the massive stars.

{\it LH9:}  This association was
studied in detail by Parker et al.\ (1992), 
using the same 1985 imaging data and calibration that we employ
here for many of the other associations.  The central object, HD~32228, was
clearly an unresolved cluster of many early-type stars, with a composite
WC+O spectral type.  The region was recently
examined using {\it HST}
by Walborn et al.\ (1999), and we adopt their photometry and spectroscopy
here, ignoring the region outside of the central 30 arcsec covered by the
PC frame of WFPC2.  
Although they were able to spectroscopically
observe the WC component separately from
its close neighbors for the first time, their spectral classification of WC4
is based upon only a spectrum in the blue, which lacks the crucial classification lines O~V $\lambda 5592$, C~III $\lambda 5696$
and C~IV $\lambda 5812$ (e.g., Smith 1968a; van der Hucht et al.\ 1981). 
Walborn (1977) had earlier classified the WR star as WC5, but this was also
based upon a blue spectrogram.  Smith (1968b) called the star WC5, but this was
before the earlier WC4 subclass was introduced.  Breysacher et al.\ (1999) 
cite a speckle study by Schertl et al.\ (1995) for the spectral type, but no
spectrum was actually taken as part of that study.  We adopt WC4 as the
spectral type, but note here that the type is uncertain. 
The visually brightest stars
in the LH~9 association are late-O supergiants (O9~I and O8.5~I).

{\it LH12:} Ours is the the first modern study of this association.
It contains
the WC4 star Br~10.  The visually brightest stars are B-type supergiants,
although our study has revealed a very early O-type star, with type O4~V(f).
To the extent that the association is coeval, the B-type
supergiants evolved from stars of spectral subtype O4~V or even earlier.

{\it LH31:} This association contains
two Wolf-Rayet stars, Br~16 classified by Conti \& Massey (1989) as
WN2.5.  A second WR star has been recently discovered by Morgan \& Good (1985),
who classify the star as WC5+O. This star is BAT99-20 in the
catalog of Breysacher et al.\ (1999), whose finding chart puts the star centrally located in the
association boundary shown by Lucke (1972). 
Nebulosity prevented Lucke from photographic photometry of any by the
brightest few stars. The visually
brightest stars include a B1~III, an O6~I(f), and two yellow stars.  One
brighter of these, which we call LH31-1002, is apparently an LMC F2 supergiant,
based both upon our measured radial velocity and strong Sr~II $\lambda 4077$
(see Jaschek \& Jaschek 1990).  The other is clearly a late F-type foreground
dwarf, based upon its radial velocity and its lack of Sr~II.

{\it LH39:} The cluster was examined by Schild (1987), and again by
Heydari-Malayeri et al.\ (1997).  We obtained new photometry and a few
additional spectral types.  The association contains one of the rare
Ofpe/WN9 stars, Br~18=Sk$-69^\circ79$.  Ardeberg et al.\ (1972) list
the star Sk$-69^\circ80$ as having a spectral type of F2~Ia; however,
Schild (1987) suggests a type of B8:~I.  Our photometry is
consistent with something intermediate between these two, and we will use
its photometry to place it in the HR diagram.  (The radial velocity of
Ardeberg et al.\ does confirm it is an LMC member.)   
We will find that two A supergiants classified by Schild (1987) appear to be
much older than the rest of the cluster.  We have independent spectroscopy for
one of these, LH39-22, and confirm Schild's type.

{\it LH41:} This association contains S Doradus, the prototype LBV,
and the visually brightest star in the cluster.  The second brightest
star, R~85=Sk-69$-69^\circ92$ we propose as an LBV, based upon its 
spectral and photometric variability, as discussed earlier in 
Section~\ref{Sec-r85}.
The third brightest star is the Wolf-Rayet
star Br~21, classified by Conti \& Massey (1989) as B1Ia + WN3.
The star LH41-4 is of M-type, but we lack the radial velocity information
that would ascertain whether this is an M supergiant or foreground dwarf.
There are two lower luminosity but bona fide A-type supergiants, and an
F5 supergiant.  The latter has been confirmed based upon our radial velocity
and the strength of Sr~II $\lambda 4077$. (It is also an excellent match
to the F5Iab star HD~9973 shown in the Jacoby, Hunter, \& Christian 1984 atlas.)
Ours is the first modern study of this association.

{\it LH43:}  The visually brightest star is an early
M-type, but again we lack the proper radial velocity information to
ascertain whether this is an LMC member or not.  The second brightest
star is a newly discovered O4~If star.  The WR star Br23 is classified
WN3.

{\it LH47:} This association was studied by Oey \& Massey (1995) and
Will, Bomans \& Dieball (1997).  We adopt the photometry and spectroscopy
of the former, who obtained spectral types for all the brighter components,
primarily of early to mid O-type.  Oey \& Massey (1995) suggest that there
are two ages for the stars in the LH47/48 region: stars interior to the
DEM~152 superbubble have an older age than stars in rim of the bubble.
The WR star and other
massive stars of interest are on the exterior, and we will restrict
our analysis to those.
In agreement with Will et al.\ we find no difference between the
photometric $Q$ and that expected on the basis of spectroscopy;
we cannot comment on their
assertion that field-to-field differences exist in the individual $B-V$
and $U-B$ colors at the 0.15 mag level, other than to note our value for
the reddening appears to be reasonable.  

{\it LH58:} This association was recently studied by Garmany et al.\ (1994).
It contains three WR stars, Br~32 (WC4+abs),
Br~33 (WN3+abs), and Br~34 (B3I+WN3). The latter is the visually brightest
star.   We did obtain a spectrum of the earliest-type star in the association,
reclassifying it from O3-4~V to O5.5~V((f)), as described earlier
(Section~\ref{Sec-Os}).
We note that LH58-473 as B0.5V must be a giant based upon its
M$_V$.

{\it LH64:} This association was studied by Westerlund (1961) as well
as by Lucke (1972).  Ours is the first modern study.  The three visually
brightest stars have colors characteristic of mid-to-late type stars,
presumably foreground, although spectroscopy is needed to
determine if they are supergiants.  The WR star Br~39 was not classified
by Conti \& Massey (1989), but was called WN3 by Breysacher (1981).

{\it LH81:} Also studied by Westerlund (1961) and Lucke (1972), ours is
the first CCD study of this interesting region.  It contains
three WR stars: the WC4 star Br~50 (classified by Conti \&  Massey
(1989), the WN4+OB star Br~53 (classified by Breysacher 1981), and
Sk$-69^\circ$194, discovered as a WR star here (B0I+WN).  The visually
brightest star is a foreground G dwarf.  We identify two very
early-type stars in the association, W28-23, a O3~V((f)) star, and
W28-5.  As discussed in Section~\ref{Sec-o3s}, we classify the latter as O4~V((f)) based upon its He~I to He~II
strengths, but our very high S/N spectrum 
shows the definite presence
of N~V $\lambda 4603,19$ absorption lines, previously associated only
with O3 stars.  Possibly an intermediate type (O3.5) would be
warranted, but we leave that until we have been able to complete a
detailed analysis of this star.

{\it LH85:} We identify the star LH85-10 as a newly discovered
B[e]. Our study is the first since Westerlund (1961)
and Lucke (1972).  The association also contains the WR star Br~63,
classified as WN4.5 (Breysacher et al.\ 1999).
Westerlund (1961) treated this association and the neighboring LH~89
as one unit; we treat them separately here, following Lucke (1972), although
the ages and cut-off masses we derive will prove to be essentially the same.
The earliest spectral type we find in LH~85 is 
B0.5.

{\it LH89:} A section of LH89 was included in the study by
Schild \& Testor (1992) of stars in the general 30 Doradus region
(their ``zone 3"), in addition to the Westerlund (1961) and Lucke (1972)
studies.  We have used their spectral types as a supplement to our own,
but use our own CCD photometry. 
The association contains Br6 (WN4) and
Br~64=BE~381, the archetype of Ofpe/WN9 stars.
The visually brightest stars are three tenth magnitude 
stars of intermediate color; radial velocities
of the two brightest demonstrate that they
are LMC members (Ardenberg et al.\ 1972).  Our spectrum of the third shows
it is a foreground F8 dwarf, based both on its radial velocity
and the weakness of high-luminosity features in the spectrum,
emphasizing once again the need for
spectroscopy in determining membership of even bright stars in the Clouds.
We will find that the
two confirmed A-F supergiants turn out to be coeval with the rest of
the association members.

{\it LH90:} Photometry of the LH~90 region was published by Schild \& Testor
(1992), who refer to the region as ``Zone 2", and provide a finding
chart in their Figure~3.  (Only stars 2-33, 2-34, and and 2-45 fall
outside the association boundary shown by Lucke 1972.)  There are three
clumps of stars, designated as ``clusters" $\alpha$, $\beta$, and
$\delta$ by Loret \& Testor (1984).  The region was re-examined by
Testor et al.\ (1993), who provided new photometric and
spectroscopic data on knots $\alpha$ and $\beta$.
Clusters $\beta$ and $\delta$ were also studied by
Heydari-Malayeri et al. (1993).
Recently, Walborn et
al.\ (1999) were largely successful in further unraveling the
$\beta$ knot of stars using
WFPC-1 images and FOS spectroscopy with {\it HST}. (They refer to
$\beta$ alternatively as ``NGC~2044 West" and ``HDE 269828".)  To this,
we add our own {\it UBV} photometry and spectroscopy.  
We note that a comparison of the high resolution image of Testor et al.\
(his Fig.~1b) with that of Walborn et al.\ (Fig.~5) suggest that 
ground-based
work actually did a remarkably good job of resolving multiple components in
cluster $\beta$. The stars designated ``TSWR2" and ``TSWR1" are multiple,
but the others are actually well resolved with 1" resolution.
The components found independently by our PSF-fitting are an exact
match to those identified by Testor et al.
The most interesting star is the one Testor et al.\ identify as ``6" in
cluster $\beta$; this is the star labeled ``9" by Heydari-Malayeri et al.,
and split into two components (``9A" and ``9B") by Walborn et al., although
9B is 1.5 mag fainter than 9A and hence the composite spectrum we obtained
from the ground is a good representation of star 9A.
We have noted earlier (Section~\ref{Sec-o3s}) that the star $\beta-13$ is
probably better considered an O3~If* star rather than the O4If+ used by
Walborn et al.

In our analysis of this region we will make use of our new ground-based data,
but defer to the {\it HST} data of Walborn et al.\ for stars for the
the group of stars called ``TSWR1" (or $\beta$-6) by Testor et al.,
which is the star identified as
``5" by Heydari-Malayeri et al., split into multiple components by
Walborn et al. (1999).
Our ground-based (composite) spectrum would have resulted in
a ``B0I+WN" designation, but the {\it HST} work clearly shows that
these are separate stars, in accord with Testor et al.'s finding.
One wonders if other ``BI+WN" systems might
not be similarly resolved.  
We also note the need for a high-resolution study of the $\delta$
knot in this interesting region.

In addition to the WN4 component of ``TSWR1", the association contains
many other WRs:
Br~56 (WN6), Br~57 (WN7), Br~58 (WN5-6), and Br~65 (WN7), all of fairly
late type for the LMC, plus the WC4 star Br~62.   The classifications are
from Conti \& Massey (1989), except for that of Br~65, which is from 
Breysacher (1981).  Earlier (Section~\ref{Sec-wrs}) we suggest 
that Br~58 may be better classified as O3If*/WN6.

In analyzing this cluster in Section~\ref{Sec-coeval}, we find that the
$\beta$ subclump is no more coeval than the association
as a whole, as witness the fact that both a B0~I star of modest
luminosity cohabits with an O3 star of high luminosity.
There is a significant range of ages.

{\it LH101:} This region has recent CCD photometry and spectroscopy by 
Testor \& Niemela (1998).  To this, we obtained our own spectra for
three of the stars, as discussed in Section~\ref{Sec-spectra}.  
We find that ST5-27 is an O3V((f)), as indicated both by
the lack of He~I and the weak presence of N~V $\lambda 4609,19$ absorption; 
the star was classified as O4~V
by Testor \& Niemela.  We confirm that their star ST5-31 is indeed an O3If.
And, we reclassify ST5-52 as an ON5.5V((f)) star, rather than O3~V (Section~\ref{Sec-Os}).
The association contains Br~91, another of the rare Ofpe/WN9 objects.

{\it LH104:} This association was also studied by Testor \& Niemela (1998).
We have obtained new CCD photometry, as well as additional spectroscopy.
The association contains three WRs, all of which are spectrum binaries
as described by Testor \& Niemela: Br~94 (WC5+O7), Br~95 (WN3+O7), and
Br~95a (WC5+O6).  The visually brightest star is the  B[e] star, S~134
(Zickgraf 1993).  We note that one of the visually brighter stars is an M
star, confirmed by Testor \& Niemela as a supergiant on the basis of its
radial velocity; this agrees with the conclusion of
Massey \& Johnson (1998)
that WRs and M supergiants are sometimes found in the same associations, contrary
to the prevailing wisdom.

\section{Construction of HRDs: 
Coevality and Uncovering the Most Massive Stars}
\label{Sec-hrd}

In order to identify the most massive stars, we construct ``physical"
H-R diagrams ($\log T_{\rm eff}$ vs. $M_{bol}$) for comparison with 
the theoretical evolutionary tracks.  These tracks will allow us to 
test for coevality, and determine the masses for the
highest mass unevolved (H-burning) stars in these associations.
First, we must correct the observed photometry for reddening, and second
to convert the data (spectral types and photometry) to effective temperatures
and bolometric magnitudes.  Next we will construct the HRDs and uncover the
masses of the most massive stars.

\subsection{Corrections for Reddening and Testing the Reddening-free
Index $Q$}

Our first step in constructing HRDs is to determine the reddening corrections
for each region.  
For stars with spectral types, we adopt the intrinsic colors of FitzGerald
(1970) as a function of spectral type and compute the color excess $E(B-V)$
directly. Occasionally even a star with a spectral classification has a
reddening which differs substantially from the other members in a region,
and so we've chosen to
constrain the reddening 
to the range indicated
by the majority of stars for which there are spectral types.  We
list in Table~3 the average color excess $\overline{E(B-V)}$ and ranges of
$E(B-V)$ we adopt
for each of the 19 associations. (For consistency, we re-derived reddenings
even for the associations with values already in the literature.)

Although we obtained spectral types for most of the bright stars in each
association, there are some stars for which we have only photometry.
Rather than de-redden these using $\overline{E(B-V)}$ we 
employed a relationship between $Q$ and $(B-V)_o$ to de-redden
each star individually, using the star's photometry and $\overline{E(B-V)}$
as a gauge of whether the star's intrinsic colors were
sufficiently blue for this method to work.
We found that for stars with $Q<-0.2$ for $(B-V)_o\approx (B-V)-\overline{E(B-V)}=-0.06$ we could de-redden star by star;
for stars with intrinsic colors redder than this amount, we adopted the
average reddening. We did further constrain
the reddening to the range determined by the majority of stars
with spectral types in a region.

Since our earlier work (Massey et al.\ 1989b, 1995b) it has become clear that
the intrinsic colors as a function of spectral type or effective temperatures
are not extremely well know, particularly for the early B supergiants,
and we
have therefore computed new relationships based $Q$ and $(B-V)_o$ (and
the intrinsic colors and effective temperatures) using the 
Kurucz (1992) ATLAS9 models, using a metallicity of 0.8 times solar,
a compromise between SMC, LMC, and (local) Galactic abundances.
We find  $$(B-V)_o=-0.005 + 0.317 \times Q$$
regardless of luminosity class.

Construction of the reddening-free index $Q$ for the stars with spectral type
allows an independent check upon the accuracy of the photometry: is there
good agreement between the observed $Q$ and that $Q$ expected on the basis of
the intrinsic colors for that spectral type?  We 
determine if there is a statistically significant shift in $Q$ for all the
stars for which we have spectral types in each association. In general we find 
deviations in $Q$ within 1$\sigma$ of 0.0.  The only exceptions for our new
photometry 
are LH~43, for which we adopt a shift $\Delta Q=-0.13$, and LH~64, for
which we adopt a shift $\Delta Q=-0.15$ (i.e., in both cases the photometric
$Q$ must be made more negative to agree with the expectations of the
spectroscopy).  The two regions were imaged within a few minutes of each
other during the 1996 night at about the same time that the observing
assistant reported seeing an isolated cloud.   Interestingly, the reddening
values we found for these two regions are each quite reasonable, suggesting
that it might have been only $U$ which was affected in the two fields.
Inspection of the observing logs confirms that the {\it U} exposure of
LH~43 was observed back-to-back with the {\it U} exposure of LH~64.
The next regions observed, LH85/89, appears to have no significant
photometric problems.  We see no problems with any of the 1985 data, either
published or new in this paper.  We do find a shift of $\Delta Q=-0.11$ for
the LH~101 photometry published by Testor \& Niemela (1998).
Although the large
scatter (0.08~mag) makes this result marginal in significance, and nearly
all the stars of interest to us have spectral types, we still apply this
correction to their photometry.  

The WFPC2 photometry of LH~9 (``HD 32228") by Walborn
et al.\ (1999) also shows a systematic shift in $Q$, with 
$\Delta Q=-0.07\pm0.01$(s.d.m.)~mag.
Presumably this shift is an artifact of their reduction procedure. 
This shift is larger
than any of the ground-based {\it UBV} data reported here, other than the
cases noted above, and so it is unlikely due to any problems with the
spectral-class to $Q$ relationship we adopt. 
We did not apply any correction to their data as
we used only the stars with spectral types in constructing the HRD, although
this could have some minor effect on the absolute magnitudes (0.2~mag) and
hence masses we determine if the problem is in $B-V$ rather than in $U-B$.

\subsection{Conversion to $\log T_{\rm eff}$ and Bolometric Luminosity}

The final step in constructing the HRDs is to use the data to determine
the effective temperature and bolometric luminosity of each star.

For stars with spectral types, we begin by adopting the spectral type
to effective temperature scale given by 
Vacca et al.\ (1996) for O-type stars, based as it is on the
results of modern hot-star models.  This will yield results that are somewhat
hotter and, thus, somewhat more luminous and massive than the older 
effective temperature scale of Chlebowski \& Garmany (1991), say, or that
of Conti (1973).
For the early B stars we were faced with a dilemma.
As discussed by Massey et al.\ (1995a) 
there is a discontinuity in the effective temperature scales of hot stars corresponding to roughly where
the modern work of Conti (1973) ended (i.e., O9.5) and earlier works took
over.  In order to smooth the transition, we have adopted the effective
temperatures of B0.5-B1 dwarfs and giants as given in Table 3-4 of Conti (1988),
as those are in excellent agreement both with what we expect on the basis
of the intrinsic colors from the model atmospheres, and with the spectral
analysis of Kilian (1992).  For B1.5 and B2 dwarfs and giants, we compromised 
between the latter two.  For the B-type supergiants, we made use of the
effective temperatures suggested by Conti (1988),
the recent spectroscopic analysis of two early B supergiants by 
McErlean, Lennon \& Dufton (1998), a comparison of the intrinsic colors
listed by FitzGerald (1970) with those of the 
Kurucz model atmospheres, and the effective temperature scale given by
Humprheys \& McElroy (1984).  In the past we have relied exclusively on
the latter; we note here though that this disagrees with the more recent
analysis by 0.1 dex from B1~I through B5~I.  It is clear that a consistent
effective temperature scale that extends from O through the B-type stars is
currently lacking, and the compromise we use here is only a stop-gap until
a comprehensive study can be done.

For stars with photometry alone, we rely upon a relationship between
the reddening-free parameter
$Q$ and $\log T_{\rm eff}$ determined from the Kurucz models; this relationship
is given in Table~4, and is appropriate for intrinsically blue stars
[($Q<-0.6$ and either $(B-V)_o < 0.00$ or $(U-B)_o<-0.6$].  For redder stars,
we use a relationship between $(B-V)_o$ and $\log T_{\rm eff}$ also given
in Table~4, based upon the Kurucz models.  The latter relationship need not
be of high accuracy, as the BC becomes a less steep function of $\log T_{\rm eff}$.

The bolometric correction (BC)
is a function primarily of effective temperature
with little dependence on $\log g$; we adopt the approximation 
$BC=27.66-6.84\times \log{T_{\rm eff}}$ appropriate to hot stars 
($\log T_{\rm eff}>4.2$) given by Vacca et al.\ (1996).
For the cooler supergiants we find discrepancies between
the BCs listed by Humphreys \& McElroy (1984) and the corresponding effective
temperatures when compared to the Kurucz models; we adopt the relationship
given in Table~4 based upon a fit of the BCs with $\log T_{\rm eff}$ based
upon the Kurucz models.

We show the resulting HRDs in Fig~\ref{fig:hrds}. In these figures, we have
indicated the stars with spectral types by filled circles, and those stars with
only photometry with open circles.  Crosses represent stars with only
photometry whose placement in
the HRD are uncertain for one reason or another: either
their transformations failed because of
unrealistic colors, resulting in superfluously high effective temperatures
and locations to the left of the ZAMS, or else
their colors are too red to allow us to
determine their reddening using $Q$, or the derived reddening falls outside
the range we adopted on the basis of our spectroscopy.  
We also mark with
an asterisk stars with spectral types but whose location is uncertain,
such as the components of double-lined binaries.
We include in these diagrams the
evolutionary tracks of Schaerer et al.\ (1993) computed at $z=0.008$
(appropriate for the LMC), and the tracks of Schaller et al.\ (1992) at $z=0.001$, similar to the $z=0.002$ of the SMC.

We also show isochrones corresponding
to ages of 2, 4, 6, 8, and 10~Myr (dashed curves), which we computed
using a program kindly provided
by Georges Meynet.  

\subsection{Identification of the Most Massive Stars, and the Limits of
Coevality}
\label{Sec-coeval}

Using the results of our calculations in the previous section, we can now
identify the mass of the highest mass unevolved (H-burning) star
in each association. We list the derived quantities ($\log T_{\rm eff}$,
$M_{\rm bol}$, mass, age) for the highest mass stars in Table~5.

For associations that are strictly coeval, we expect that the stars in the HRD
will follow a single isochrone, and in that case the highest mass would 
correspond to a ``turn-off" mass and we could be confident that any evolved
members of these associations were descended from stars with masses greater
than this value.  Alas,  the HRDs of Fig.~\ref{fig:hrds}
do not for the most part yield
such an unambiguous picture.  In all cases there is some spread across
isochrones.  If real, such spreads would tell us that the massive stars formed
over some period of time.

How significant are these age spreads?  We can answer this quantitatively by
considering the errors associated with the placement of stars in the HRD.
Let us first consider the {\it systematic} errors.
In Fig.~\ref{fig:err}(a) 
we show the location of the spectral type calibration data
in the HRD.  The huge gap among the supergiants 
(upper-most string of points) corresponds to the difference in the
adopted effective temperature of a B5~I and a B8~I star, which is a realistic
uncertainty in spectral classification. Smaller gaps
likewise correspond to differences of a single spectral type.
We have adopted an absolute magnitude corresponding to each type;
of course, our stars, with $M_V$ computed from the photometry, will fall
both above and below the points shown.  It is instructive to see the
systematic deviation of these stars from the ZAMS as one approaches cooler
temperatures among the dwarfs. By log~T$_{\rm eff}=4.2$ the locations
of the dwarfs are nearly coincident with the {\it terminal} main-sequence,
as indicated by the first switch-back in the tracks.
In this region the isochrones are
tightly spaced, and a large error in the age spread would result if we
compared the ages of a high mass luminosity class ``V" stars  with one of
lower mass; for this reason we should exclude stars below 20$\cal M_\odot$
unless they are of high {\it visual} luminosity, such as 
an A-type supergiant.

We note that this progression away from the ZAMS is intrinsic to the spectral
type to $\log T_{\rm eff}$ calibration we've adopted and/or the
absolute visual magnitude scale we've used for the purposes of this
illustration.  Transformations to effective 
temperatures on the basis of {\it colors} are usually often based on the use
of spectral types as an intermediate step, rather than going directly from
model atmosphere colors to effective temperatures.  In these cases, the
apparent presence of stars to the right of the ZAMS might be misconstrued as
evidence of pre-main-sequence objects.  
We emphasize the need for spectroscopic followups
to establish the authenticity of such discoveries. 

Next, let us consider the
{\it random} errors caused by misclassifying stars by a single spectral type
and/or major 
luminosity class; i.e., calling a star an ``O8~III" when in fact it is
an ``O9~I".  (The absolute visual magnitudes of these two subclasses overlap,
and so our photometry would pose no warning.)
We would overestimate the star's luminosity by 0.1~mag simply by assuming a
slightly too negative $(B-V)_o$, which will lead to too large a value for
$A_V$.  More significant, however, is the fact that we will
overestimate the star's 
effective temperature by 0.05~dex, and thus overestimate the star's bolometric
correction by 0.4~mag, for a net error of 0.5~mag.  The age we calculate might
be 3.80~Myr (6.58~dex) if the actual age were 5.25~Myr (6.72~dex).  We expect
misclassification by a single spectral subtype to be common.
The size of the errors we make will depend of course upon the spectral type.
We show in Fig.~\ref{fig:err}(b) 
the errors associated with misclassification of a star
by one spectral type and/or luminosity class.  (We have not included in this
figure the modest addition error caused by the change in reddening adopted;
this will increase these errors.)

Given this discussion, we can ask the question: what fraction of stars
of 20$\cal M_\odot$ and above, and lower-mass supergiants, 
are in fact consistent  with some median age for
the association?  
We assume here that our error in spectral sub-typing is only
1 subtype, except for uncertain cases. We compute the youngest and oldest ages of each star associated with such a misclassification; 
if the cluster's median age falls
within this range, we consider that the star is coeval with the rest of
the cluster.   We use only the stars for which there
are spectral information, as the errors in the HRD are much greater for
stars with only photometry. (Compare Figures~1c and 1d of Massey
et al.\ 1995b.)  
We list the fraction of stars that we find to be coeval
in Table~6, along with the median
ages of the clusters.

Even for the clusters that have a large percentage 
of stars whose ages are within 1$\sigma$ of the 
median cluster age, we might well
ask the question if the ages of the
highest mass stars are in accord with this value.
After all, we know that in some clusters intermediate mass stars form over
some period of time (several million years), 
with the highest mass stars forming over a shorter
time, e.g., NGC~6611 (Hillenbrand et al.\ 1993) and R136 (Massey \& Hunter
1998).  We include the median
age of the three highest mass stars in Table~6.  

Inspection of the HRDs in Fig.~\ref{fig:hrds}, 
and of the numbers in Table~6, suggests
that there is a natural division, and that some of these associations are
highly coeval while the coevality of the
others are more questionable. 
If the match between
the median cluster age and the age of the 3 highest mass stars is good
($<0.2$ dex, comparable to the individual errors), and a large percentage
of stars ($>80\%$) lie within 1$\sigma$ of the median cluster age, we consider
that degree of coevality is high.  Clusters that fail to meet one or the 
other criterion we consider the degree of coevality questionable.  We
consider the coevality high in 11 of our clusters, and questionable in four.
We regard the other five associations as non-coeval.  This could be
evidence that massive stars have formed over a prolonged period, possibly
with several subgroups of different ages contributing, but it may also be simply
due to line-of-sight 
contamination within the Magellanic Clouds.

The age structure
of the LH~47/48 was discussed by Oey \& Massey (1995); as mentioned earlier,
we restrict ourselves here to the stars on the periphery of the associated
superbubble, and confirm that these stars at least form a coeval unit.
LH~90 is a very interesting association located near 30~Doradus, and its
age structure was explicitly discussed by Testor et al.\ (1993),
who found ``at least" two distinct age groups (3-4~Myr and 7-8~Myr).
They attempted
to assign membership of the evolved stars to one or the other of these
populations based, not upon spatial locale, but on the basis of bolometric
luminosity, which then assumes an answer about the progenitor masses
{\it a priori}.  They found that the $\alpha$ clump itself was not coeval.
We have separately examined the $\beta$ sub-cluster using the improved data
obtained by Walborn et al.\ (1999) and find that the same age spread apparent
in the cluster as a whole is also apparent in this subclump; the $\beta$
cluster contains both a B0~I star of modest luminosity and a high luminosity
O3~If* star.  We are, therefore, forced
to abandon this very interesting region with its large number of WR stars.

We can perform one other ``reasonability test" of whether the turn-off
masses are relevant for the evolved objects.  What is the spatial separation between the three highest mass stars (which typically define the turn-off)
and the evolved objects?  We computed the projected distances, and include the
{\it median} of these three values in Table~7, which we discuss in the next
section. (We note cases where the turnoff is actually due to the binary
companion.)  Here we find that the median separation is 25~pc. As this is the
median, there is always some massive star nearer the evolved object than the
numbers shown here.  This is consistent with the notion expressed in 
Section~1.1 that coeval massive stars may have originated in the same place,
as drifts of this order are just what we expect over 3~Myr.

We can now proceed with some confidence to assign progenitor
masses to the evolved stellar content of the coeval regions.

\section{The Progenitor Masses and BCs}

\subsection{Progenitor Masses}

In Table~7 we present the main results of this investigation: what are the
progenitor masses of various evolved massive stars?  We enclose in 
parenthesis values derived from clusters whose coevality is in question,
and exclude the WR stars from the associations which are non-coeval.

What can be conclude from these values?  First, 
we find that the masses of
the progenitors of WRs in the SMC are higher than those of the LMC.  
The data are admittedly sparse, and this conclusion rests to some extent
on what mass we assign to the progenitor of AB7: the three stars with the
highest mass in Hodge~53 are all components of spectroscopic binaries.
We can be fairly certain that the progenitor mass of AV~332 was greater
than that of its companion (i.e., $>80\cal M_\odot$), although this
supposes that binary evolution itself did not play an important role in
this system.

Turning to the WRs in the LMC, we find that there is a considerable
range of progenitor masses for the WNEs, with
minimum masses of 30$\cal M_\odot$ through 60$\cal M_\odot$.  If the
more questionable cases were included this would increase the mass range.
It appears that stars covering a range of masses pass through
a WNE stage, at least at LMC metallicities.

Both of the Ofpe/WN9 stars come from associations with very low lower limits---
in fact, among the lowest in our sample.  There is a third Ofpe/WN9 star, one
located in LH~101, which also contains evolved stars of similarly low mass
(as well as higher mass evolved stars).  We might conclude then that the
Ofpe/WN9 stars in fact are not extremely high-mass stars at all, as their
association with (other) LBVs has led others to speculate.  
Our conclusion that Ofpe/WN9 stars are actually ``low-mass" (30$\cal M_\odot$)
in origin is not
new with us: St-Louis et al.\ (1998) examined five LMC associations containing
Ofpe/WN9 stars,  including LH~89 and LH~101, and suggested much the same,
although coevality was a concern for 3 of her 5 clusters.  
Schild (1987) had earlier studied LH~39, and also noticed the relative 
high age and low mass for this cluster containing an Ofpe/WN9 star.
Using the WR standard atmosphere model, Crowther et al.\ (1995a) derive
bolometric luminosities for Br~18 (R~84) and BE~381 that suggest (present)
masses of 25 $\cal M_\odot$ and 15 $\cal M_\odot$ respectively.

Three BI + WN3 stars appear in our sample. Stars with this (composite?)
type are among the brightest stars when M~33 was imaged at $\lambda 1500$
with the {\it Ultraviolet Imaging Telescope} (Massey et al.\ 1996).
To our knowledge, no BI + WN3 star has ever been demonstrated to have a
spectroscopic orbit.  We note with some interest the relatively high minimum
masses for the progenitors suggested by our study here, and we believe that
only radial velocity studies can resolve the nature of these objects.

The WCs come from high mass stars, but, interestingly,
not significantly higher than do the
WNs.  Naively this would suggest that most massive stars of mass 45-50
and above go through both a WN {\it and} a WC stage.  Similarly the WC
star in the SMC, AB8, has a high minimum mass ($>70\cal M_\odot$), not
too different from the WNs in the SMC.   

For the LBVs in the LMC and SMC we find extremely high minimum masses---among the highest of any stars in our study.  This is in accord with the
prevailing notion that they are among the highest mass stars, and owe their
photometric outbursts and dramatic spectral changes to instabilities 
inherent to high luminosity.  
The two B[e] stars in our sample have substantially different masses, in
accord with the suggestion B[e] stars come from a large range of luminosity
(Gummersbach et al.\ 1995).

Although the cluster turn-offs provide only {\it lower limits} to the masses
of the progenitors of the evolved stars, the mass functions of these and
other OB associations we've studied are generally well populated
(cf. Massey 1995a, 1995b).  Thus these cluster turn-offs should provide substantial clues to the {\it actual} masses of the progenitors.

\subsection{The Bolometric Corrections}
We next turn to computing the BCs for these evolved
stars, using the {\it observed} $M_V$ of the star, and the $M_{bol}$ of
the cluster turn-off stars.  Previous efforts to do this (cf.\ Humphreys et al.\ 1985)
relied on the fact that little change occurs in the bolometric luminosity of
a massive star as it evolves, a fact simply traced to the fact that the core
mass remains relatively unaffected during main-sequence evolution.  Here we
propose to do somewhat better, by using the evolutionary models to make
a modest correction for evolution.

Smith (1968b) introduced a narrow-band 
photometric system to reduce the
effect of WR emission lines on photometry; her {\it ``v"} filter is centered
at $\lambda 5160$ has has a zero-point tied to the system of 
spectrophotometric standards.  For a lightly reddened star with no emission,
broad-band Johnson {\it V} and Smith's 
{\it v} are equivalent. ($V-v=-0.02-0.36\times(b-v)$ according to Conti
\& Smith 1972; a typical $b-v$ value
for a MC WR star is -0.1~mag, e.g. Table~VI of Smith 1968b. See also
Turner 1982.)  We therefore
use the {\it ``v"} mags listed by Breysacher et al.\ (1999)
when available to compute $M_V$, using the average reddenings we find
for each association.  We list these values in Table~7.

We can make two assumptions for computing the BCs.  The first
of these is to assume that the bolometric luminosity of the WR star is the
same as that of the cluster turn-off.  The second is to attempt to make a
correction for the luminosity evolution that the models predict.   The 
difficulty with the latter is that what the evolutionary models predict 
is a very sensitive function of how mass-loss is treated, and, as we
emphasized earlier in this paper, the episodic shedding of mass during the
LBV phase can play an appreciable role and is difficult to model.  The
Geneva models do not produce WR stars when standard mass-loss rates are
applied except at the very highest masses, and for this reason mass-loss
rates twice that of the observed values have been assumed in some of the
model calculations (e.g., Meynet et al.\ 1994).  From the end of core H-burning
(similar to the stage of the highest mass stars near the cluster turn-off) to
the end of the WR phase, the evolution amounts to -1.1~mag to +0.5~mag
at LMC metallicities, and +0.1~mag to +0.2~mag at SMC metallicities in
the sense of $M_{\rm bol}$ at the end of core H-burning {\it minus}
$M_{\rm bol}$ at the end of stellar models. 
We include the BCs in Table~7 computed both ways, using the $M_{\rm bol}$
corresponding to the end of core-H burning (i.e., the terminal age
main-sequence, or TAMS) and corresponding to the adopted mass of the cluster
turn-off.  

We see that the BCs for the WNE stars are indeed very negative, approximately
$-6$~mag, whether evolution is taken into account or not.  This is in good
accord with similar analysis of Galactic clusters by Humphreys et al.\ (1985)
and Smith et al.\ (1994), although this is considerably more negative than 
that of even
the earliest O-type stars ($-5$~mag).  However, recent applications
of the ``standard WR model" applied to ``weak-lined"  
WNE stars
by Crowther et al. (1995c) have found similar values for the BCs, giving us
confidence both in our method, and providing yet another indication that
the models provide a solid basis for interpreting the spectra of WR stars.
There is a large range present for the BCs of WNE stars shown in Table~7,
with perhaps some trend with spectral subclass; i.e., more negative with
earlier type.  It will be interesting to see if additional atmosphere
analysis produces similar results when applied to WN2 stars.

The Ofpe/WN9 stars have far more modest BCs ($-2$ to $-4$~mag); analysis
by Crowther et al.\ (1995a)
of Br~18 (R~84) BE~381 using the ``standard WR model" derives BCs of $-2.6$ and
$-2.7$~mag, also in good agreement with what we find.

Turning to the WCs, we find BCs of order $-5.5$~mag.  This is a little more
negative than what Humphreys et al.\ (1995) and Smith et al.\ (1994) found,
although none of the WCs in their samples were as early as those studied
here.

The BCs for S~Dor and R~85 are very modest ($-2$~mag).  
Crowther (1997) computes a similar BC for the LBV R~127, although we note
that this star is another Ofpe/WN9, or was until its outburst. We have used
our own photometry obtained of HD~5980 obtained in 1985 (Massey et al.\ 1989b)
to compute its absolute visual magnitude; given the complicated nature of
this (multiple) star, it is unclear what to make if its value.  The bolometric
luminosity of S~134 computed by Zickgraf et al.\ (1986) is $\sim -10$,
in excellent agreement with the assumptions here.

\section{Conclusions, Discussion, and Summary}

Our photometric and spectroscopic investigation of 19 OB associations in the
Magellanic Clouds has found that most of the massive stars have
formed within a short time ($<$1~Myr) in about half of the regions
in our sample. Their degree of coevality is similar to that found
by Hillenbrand et al.\ (1993) for NGC~6611, i.e., that the data are
{\it consistent} with all of the massive stars ``having been born on a
particular Tuesday." In other regions,
star-formation of the massive stars may have proceeded over a longer time,
as suggested by
 the presence of evolved stars of 15-20$\cal M_\odot$ (suggesting
ages of 10~Myr) along with unevolved stars of high mass (60 $\cal M_\odot$)
with ages of only 2~Myr.  In some cases such apparent non-coevality may be due
to chance line-of-sight coincidences within the Clouds, or to drift of lower
mass stars into the space occupied by a truly coeval OB association, but
in other cases, such as the $\beta$ subcluster of LH~90, one is forced to
conclude that star-formation itself was not very coeval, but proceeded over 
several million years.

The turn-off masses of the coeval associations have provided considerable
insight into the evolution of massive stars.  We find that only the
highest mass stars ($>70 \cal M_\odot$) become WRs in the SMC.
The numbers are admittedly sparse, and an additional
complication is the fact that most SMC WRs show the presence of absorption
lines. Are these absorption lines indicative of a weak stellar wind 
(as evidenced by the weakness
of the WR emission lines) or are these all due to binary companions?  
Conti et al.\ (1989) discuss this without reaching any conclusions, and we note
here that the issue of the binary frequency of the SMC WR stars requires
further
investigation.  Possibly a strong stellar wind due to very high luminosity
{\it and}
binary-induced mass-loss is needed to become a WR star in the low metallicity
of the SMC.

In the LMC the mass limit for becoming a WR star would appear to be a great deal
lower, possibly 30$\cal M_\odot$.  Stars with a large range of initial masses
(30-60 $\cal M_\odot$), and possibly {\it all} massive stars with a mass
above 30$\cal M_\odot$ go through a WNE stage in the LMC.  Most WR stars in the
LMC are of early WN type; this is not true at the higher metallicity of the
Milky Way, where WN3 and WN4 stars are relatively rare.  This is consistent
with recent theoretical work 
of Crowther (2000), who finds that varying only
the abundance in synthetic WN models 
(holding all other physical parameters consist) changes the spectral subtype,
with WNEs characteristic of low abundances, and WNLs characteristic of higher
abundances.  Thus, it may be the excitation classes are related neither to
the masses nor to stellar temperatures.

The true LBVs 
occurs in clusters with very high turn-off masses ($\approx 85\cal M_\odot$),
both in the LMC and the SMC.  This is very similar to the turn-off mass in
the Trumpler 14/16 complex with which the Galactic LBV $\eta$~Car is associated
(Massey \& Johnson 1993).  This supports the standard picture, that LBVs are
an important, if short-lived, phase in the evolution of the most massive stars,
at least at the metallicities that characterize the Magellanic Clouds and the
Milky Way.  We note with interest the important study by King, Gallagher, \& Walterbos (2000), who find that some LBV stars in M~31 appear to be found in
relative isolation, leading them to question whether these are all high
mass stars, at least at the higher metallicity of M~31.

The Ofpe/WN9 stars, some of which go through some sort of outburst, cannot
be ``true" LBVs, if the nature of the latter is tied to extremely high
bolometric luminosities.  We find that the Ofpe/WN9 stars have the {\it lowest}
masses of {\it any} WRs, with the progenitors possibly as low as 25$\cal M_\odot$.    Similarly, the connection of the B[e] stars to LBVs seems tenuous
on the basis of mass or bolometric luminosities.

We know that the relative
number of WC and WN stars change drastically throughout the Local Group, in
a manner well-correlated with metallicity (Massey \& Johnson 1998).  
One obvious
interpretation of this is that it is much harder to lose enough mass to
become a WC star in a low-metallicity environment; i.e., only the most
luminous and massive stars have sufficiently high mass-loss rates to achieve
this.  And, similarly, the limit for WN stars should be higher in lower
metallicity systems.  As long as the bar is somewhat lower for achieving WN
status compared to WC status, then the IMF assures that the WC/WN ratio will
change.  Thus our finding here that WCs and WNs come from similar mass
ranges (although higher in the SMC than in the LMC), suggest that an
alternative explanation is needed.
Instead, it may be that it is the relative lifetimes of the WC and
WN stages which are different at different masses; i.e.,
at very high masses the WC stage is shorter compared to the length of the WN
stage than at lower masses.  Or, it could be that the metallicity itself
affects the relative lifetimes of the WC and WN stages.  We note that we
found luminous red supergiants (RSGs) cohabiting with both WNs and WCs in many
OB associations in more distant galaxies of the Local Group (Massey \& Johnson
1998; see for example their
Figs.~14-16).  While we were unable to evaluate the degree of coevality
of these associations, the statistics suggest that these stars
have similar progenitor mass at a given metallicity, 
and that variations in the relative number of
RSGs to WRs are due primarily to changes in the relative lifetimes due to
the effect of metallicity on the mass-loss rates (Azzopardi, Lequeux, 
\& Maeder 1988).

We conclude that the BCs of WNE stars are quite substantial,
$-6$ mag.  This value is in very good accord with that determined from
weak-lined WNE stars using the WR ``standard model" of Hillier (1987, 1990)
by Crowther et al.\ (1995c). The earliest-type
WN star known (of type WN2) 
is included in our sample, and our data suggest an even
more striking BC ($<-7.5$~mag); a full analysis of Br~4 via the standard model
would be of great interest.
For the Ofpe/WN9 stars we find BCs of $-2$ to
$-4$~mag, again in good agreement with the atmospheric analysis of several
such stars by Crowther et al.\ (1995a).  We find here that the BCs of WC4
stars are typically about $-5.5$~mag.

In the next paper, we will extend this study to the higher metallicities found
in our own Milky Way galaxy.

\acknowledgements

We are grateful to Nichole King for correspondence on the issue of LBVs
and their native environments, as well as useful comments on the manuscript.
Deidre Hunter was also kind enough to provide a critical reading of
the paper.
We thank Bruce Elmegreen for correspondence and helpful preprints concerning coevality in extended regions.  Comments by an anonymous referee resulted in
improved discussion.
Classification of some of the older spectra
were done in collaboration with C. D. Garmany.
Bruce Bohannan kindly allowed us to use his
photographic spectrum of R~85 in this work. The participation of one of the authors (E.W.) was made possible through
the Research Experiences for Undergraduate Program, which 
was supported by the National Science Foundation under Grant No. 9423921.
P.M. acknowledges the excellent support provided by the CTIO TELOPS group.

\clearpage
\begin{figure}
\epsscale{1.15}
\plotone{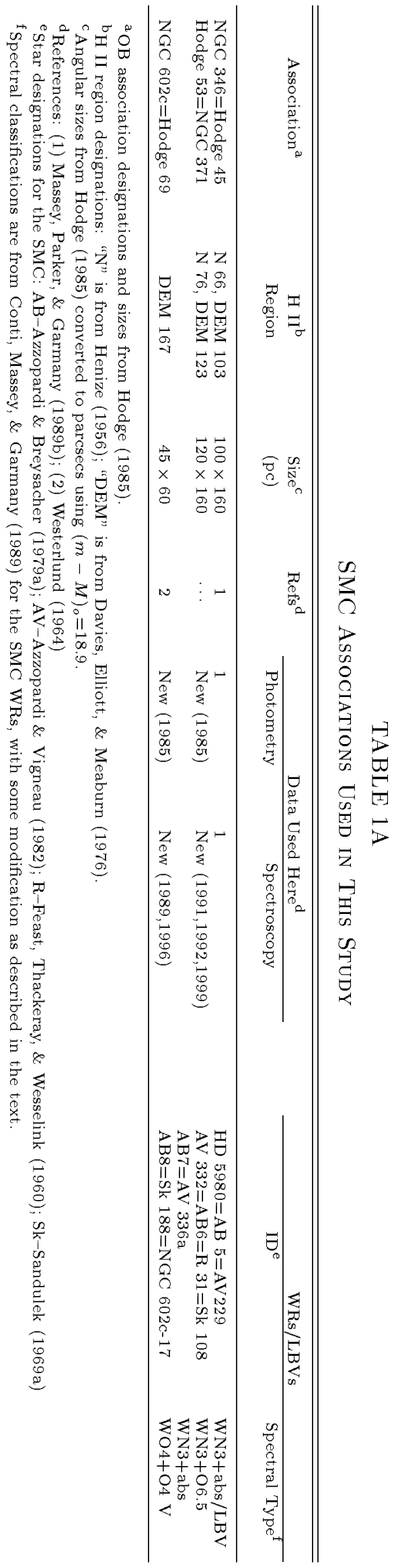}
\end{figure}

\clearpage
\begin{figure}
\epsscale{1.15}
\plotone{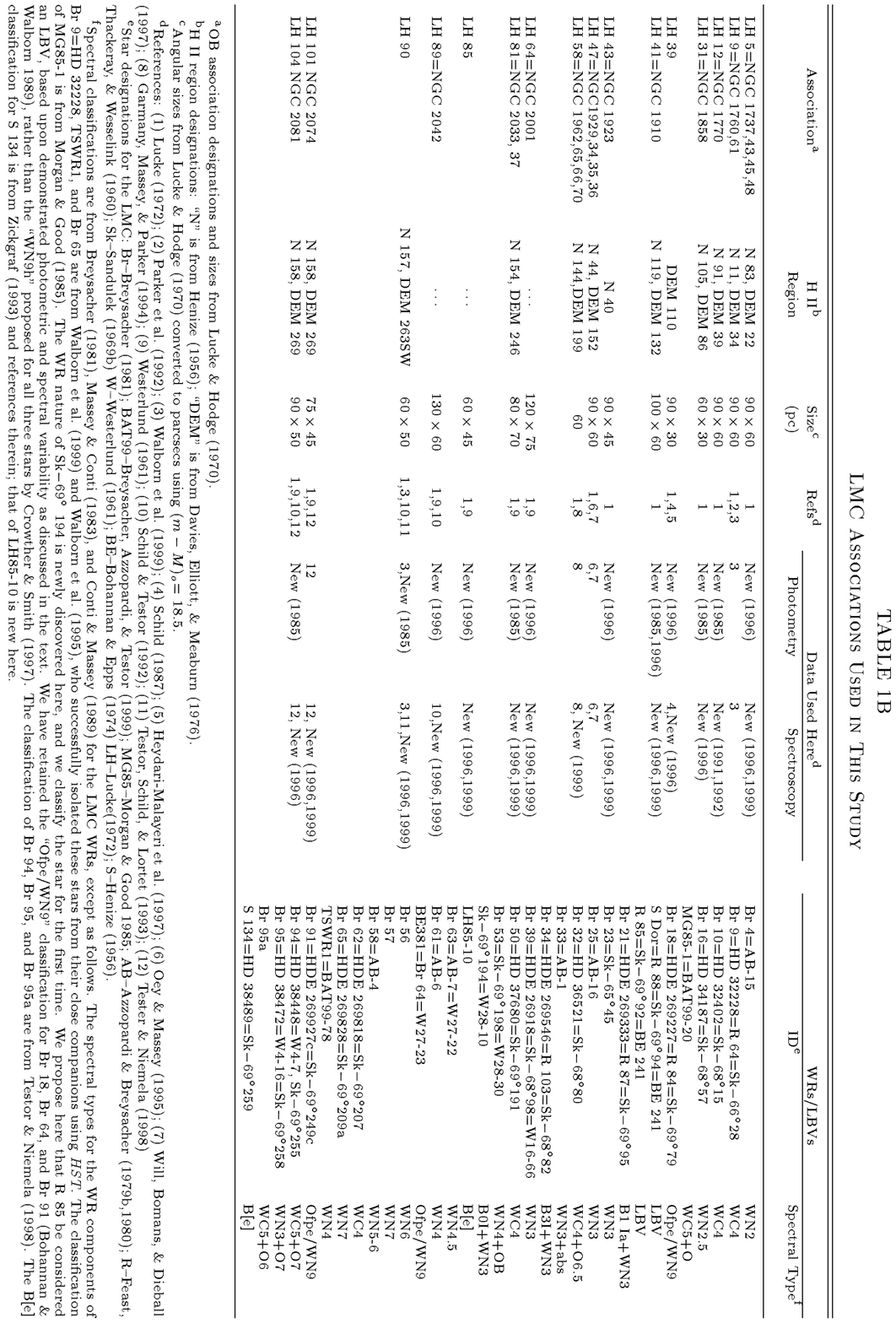}
\end{figure}

\clearpage
\begin{figure}
\epsscale{1.15}
\plotone{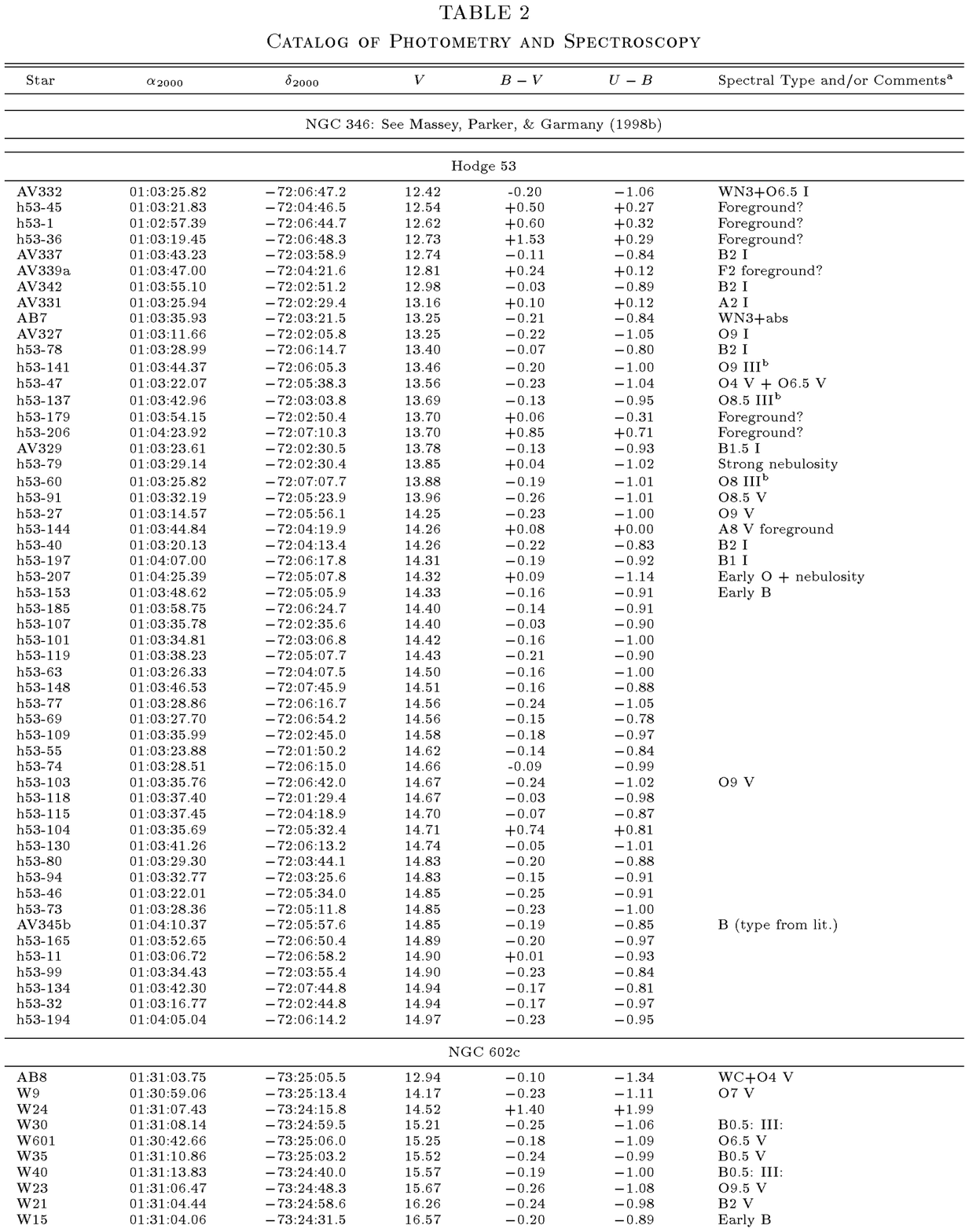}
\end{figure}

\clearpage
\begin{figure}
\epsscale{1.15}
\plotone{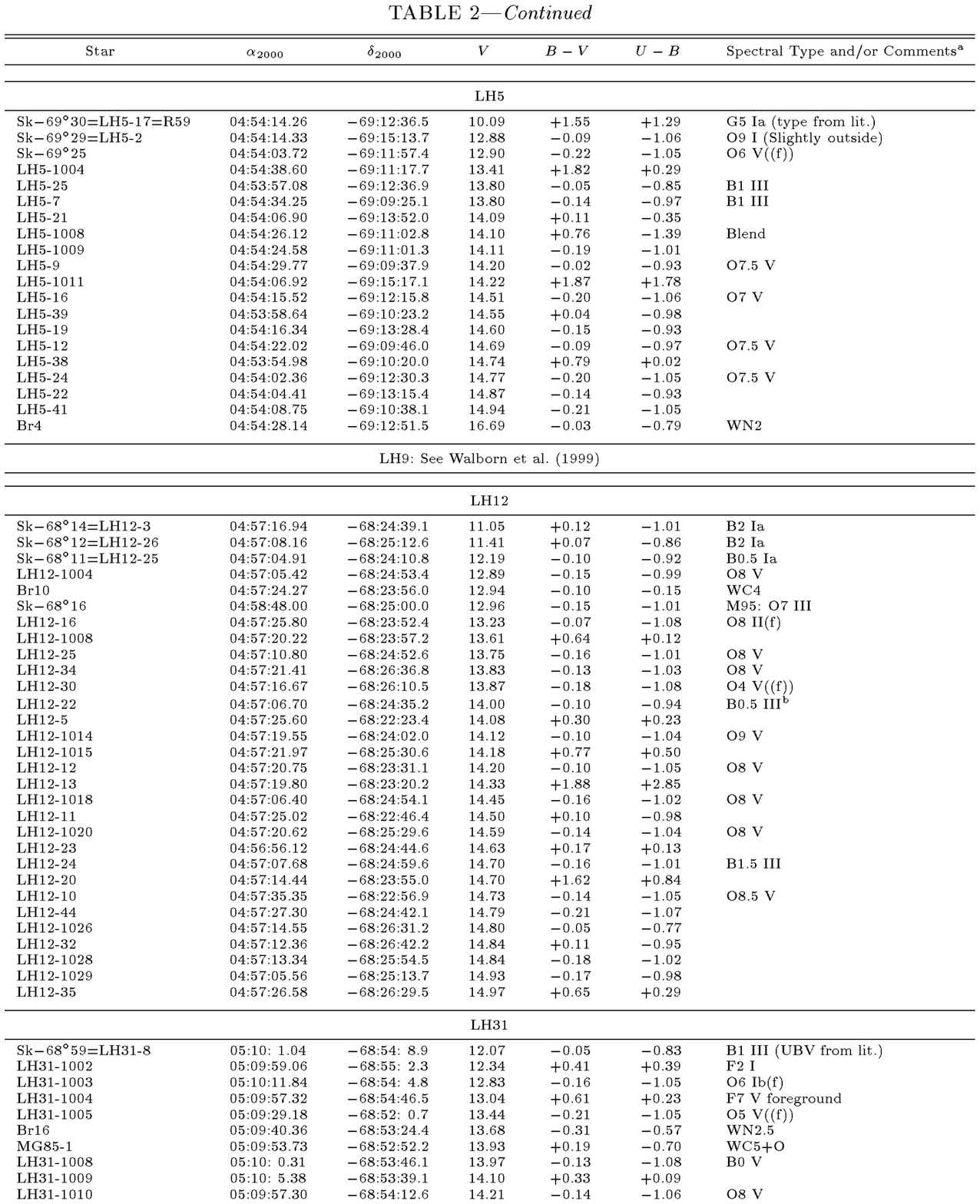}
\end{figure}

\clearpage
\begin{figure}
\epsscale{1.15}
\plotone{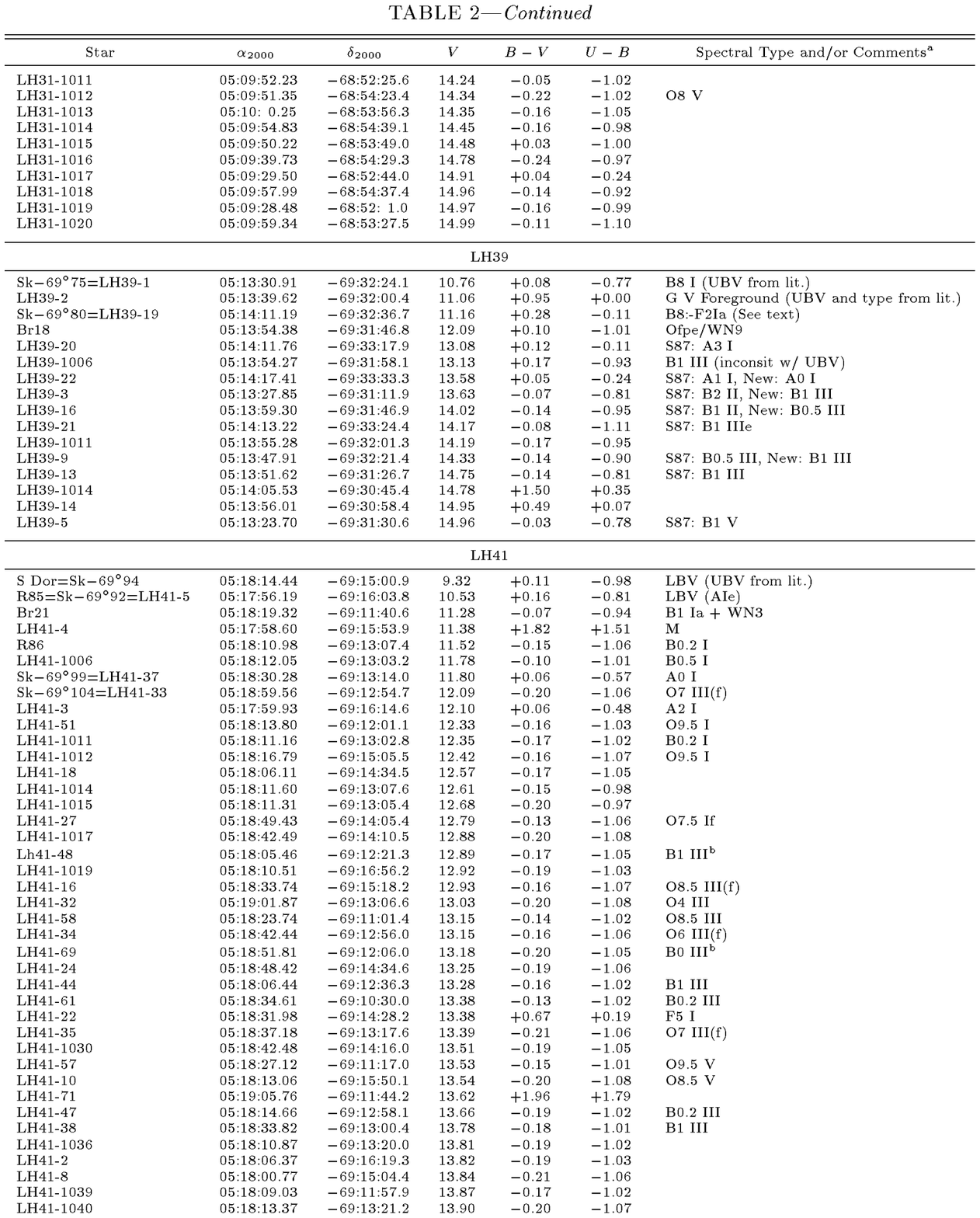}
\end{figure}

\clearpage
\begin{figure}
\epsscale{1.15}
\plotone{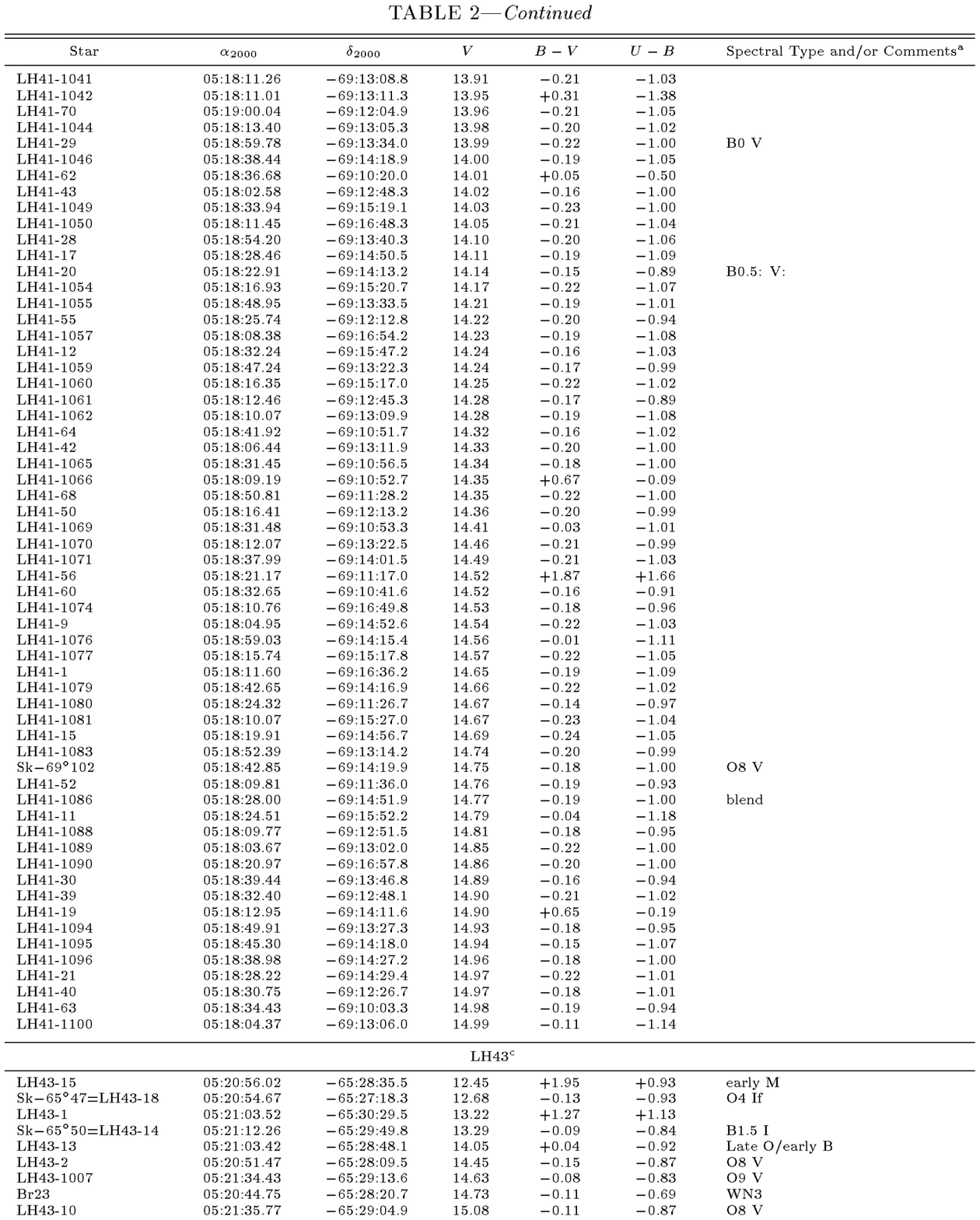}
\end{figure}

\clearpage
\begin{figure}
\epsscale{1.15}
\plotone{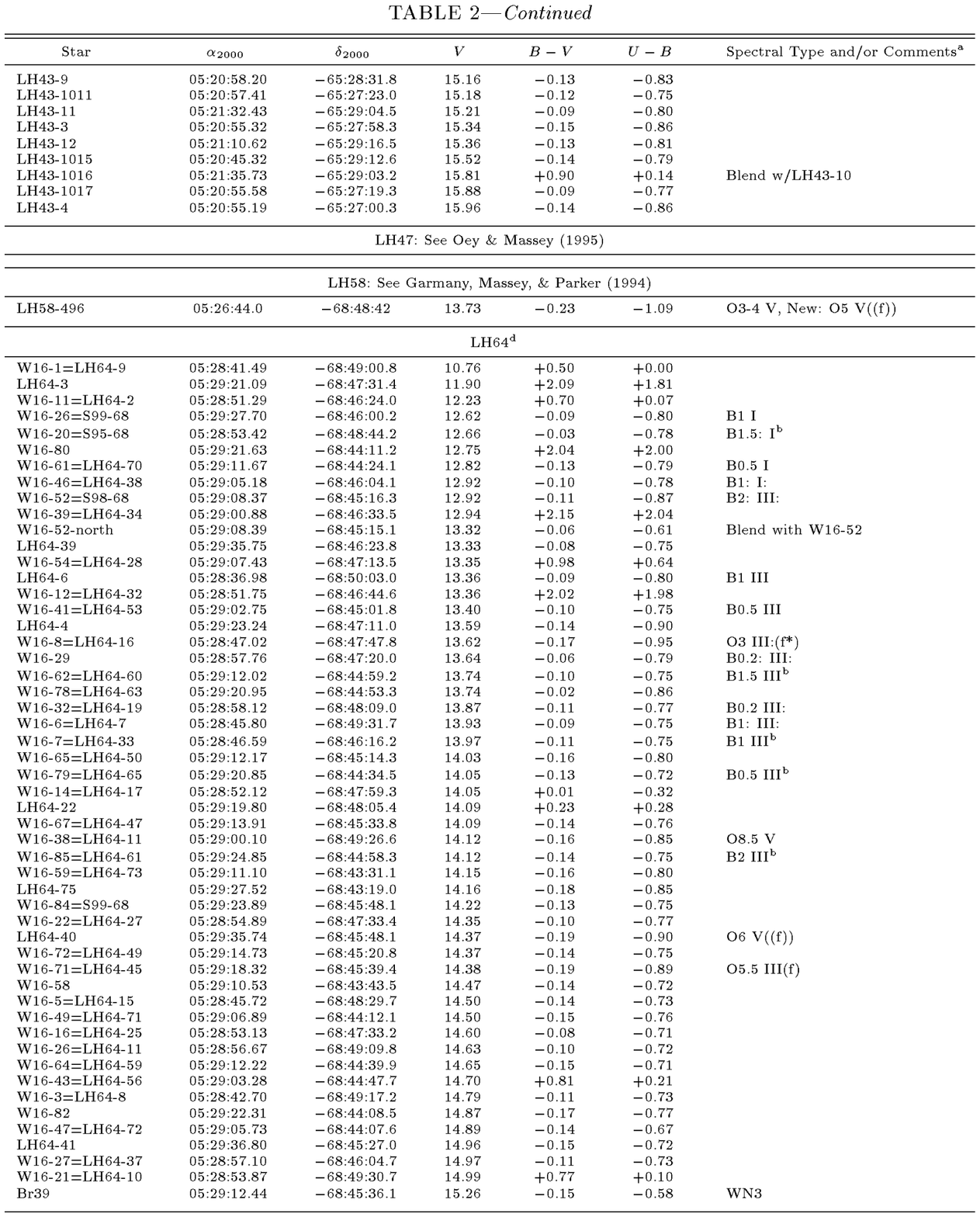}
\end{figure}

\clearpage
\begin{figure}
\epsscale{1.15}
\plotone{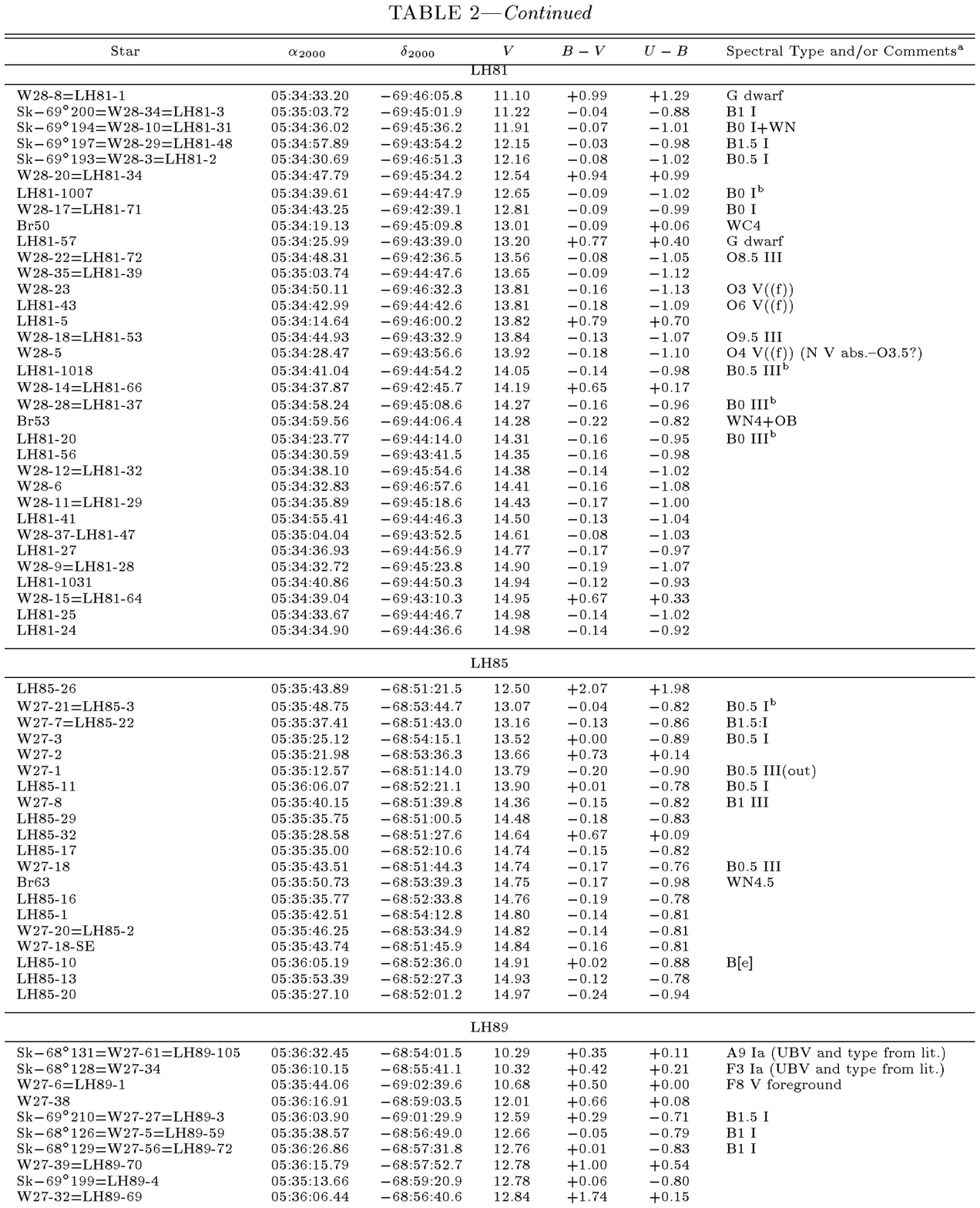}
\end{figure}

\clearpage
\begin{figure}
\epsscale{1.15}
\plotone{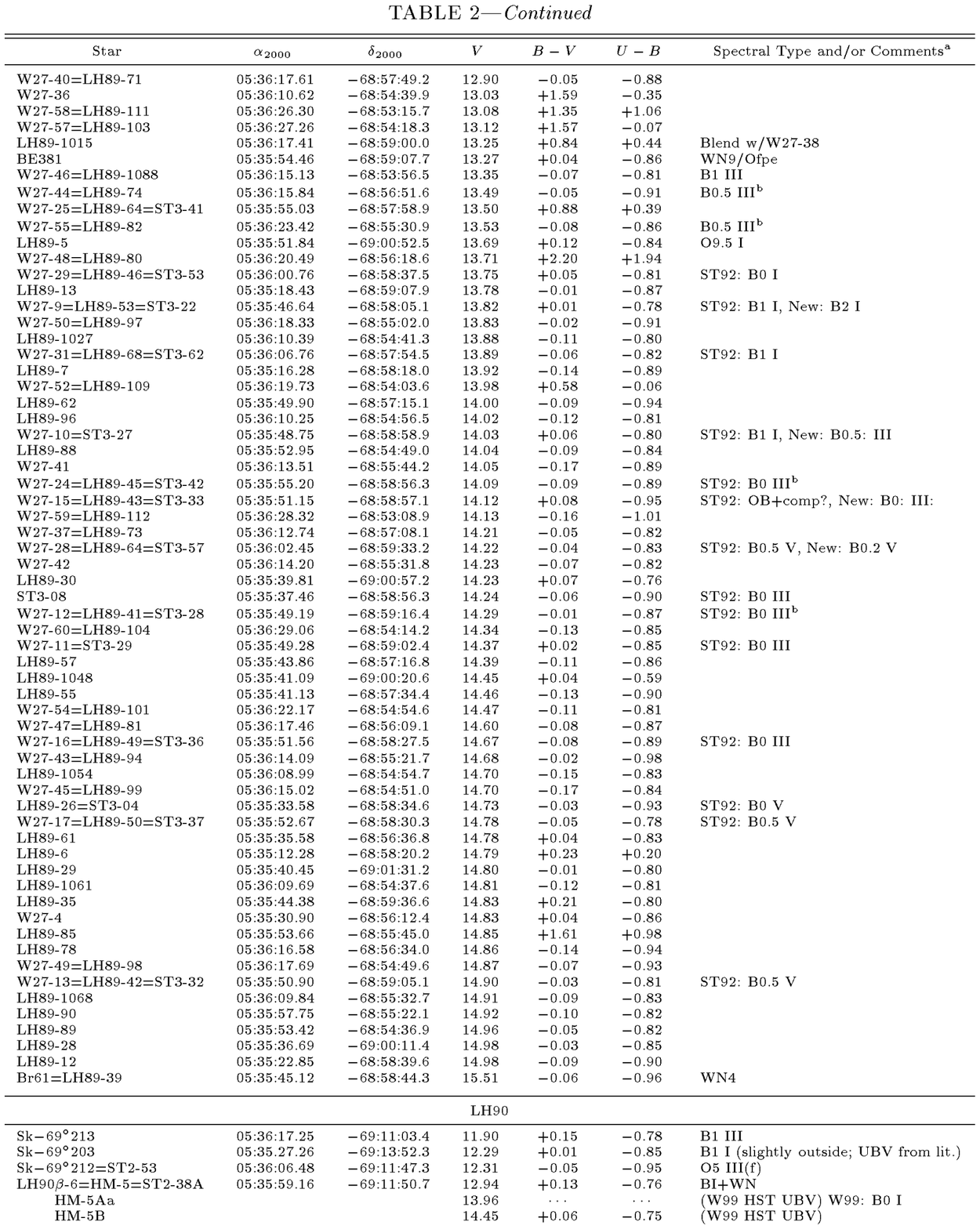}
\end{figure}

\clearpage
\begin{figure}
\epsscale{1.15}
\plotone{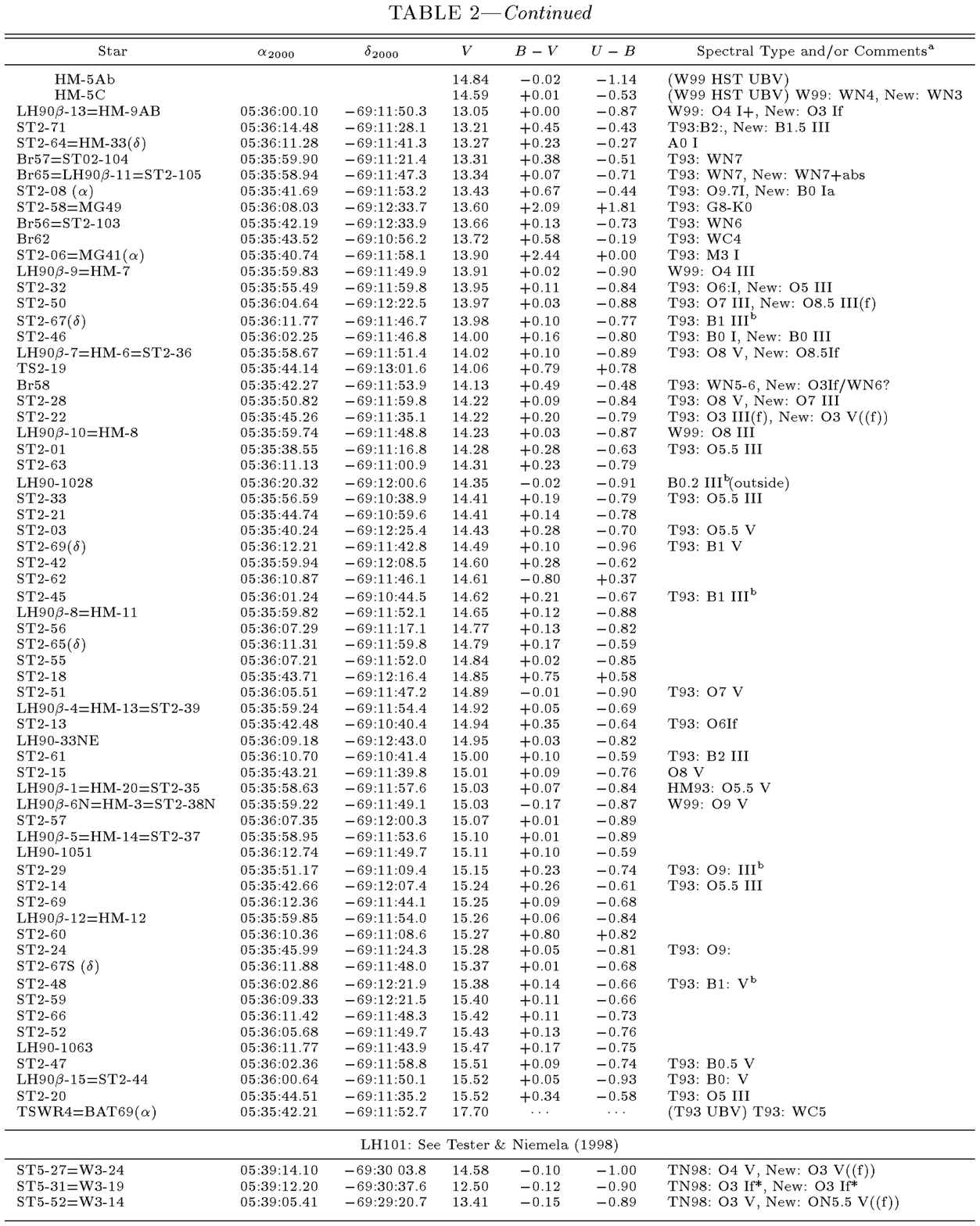}
\end{figure}

\clearpage
\begin{figure}
\epsscale{1.15}
\plotone{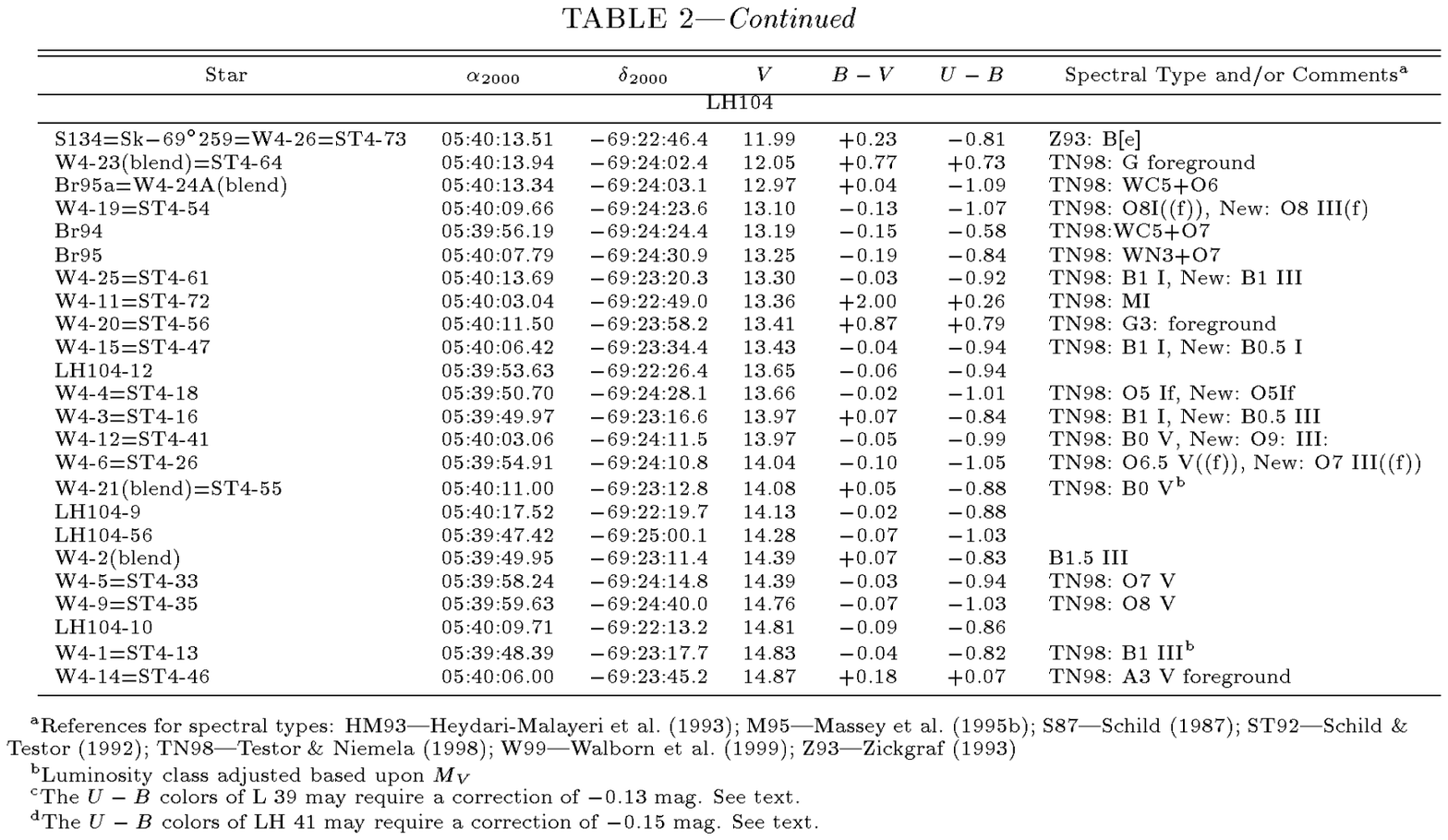}
\end{figure}

\clearpage
\begin{figure}
\epsscale{1.15}
\plotone{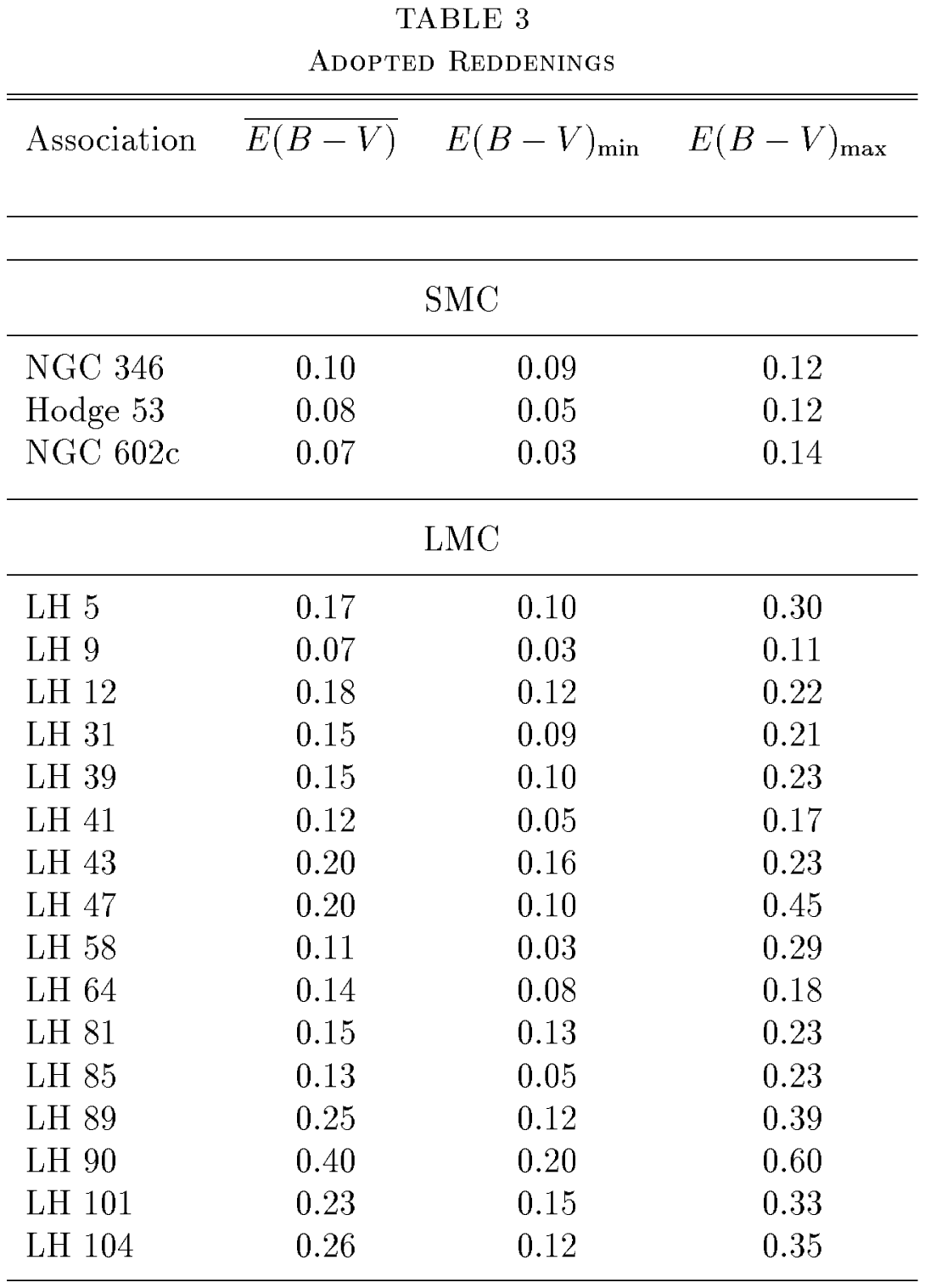}
\end{figure}

\clearpage
\begin{figure}
\epsscale{1.15}
\plotone{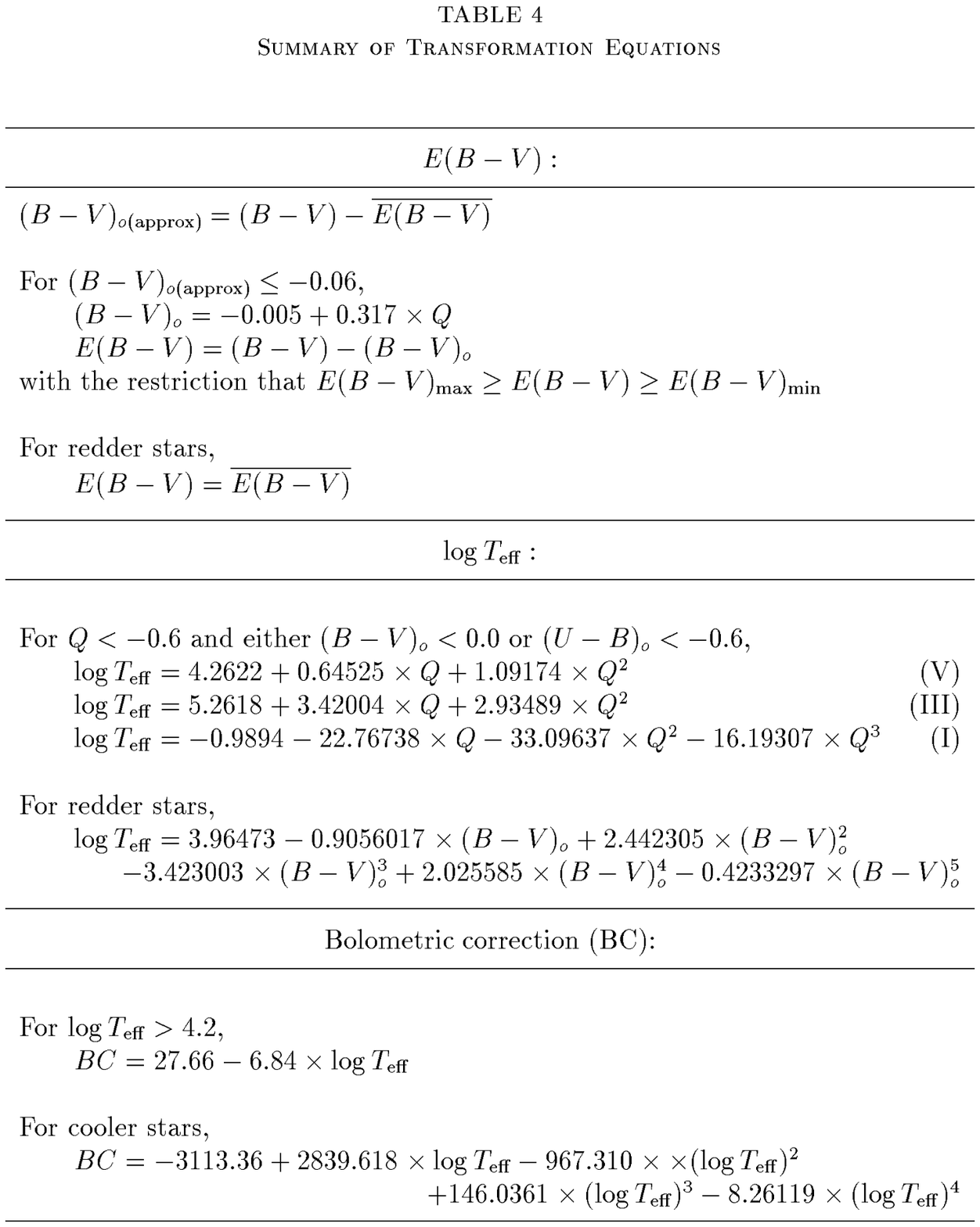}
\end{figure}

\clearpage
\begin{figure}
\epsscale{1.15}
\plotone{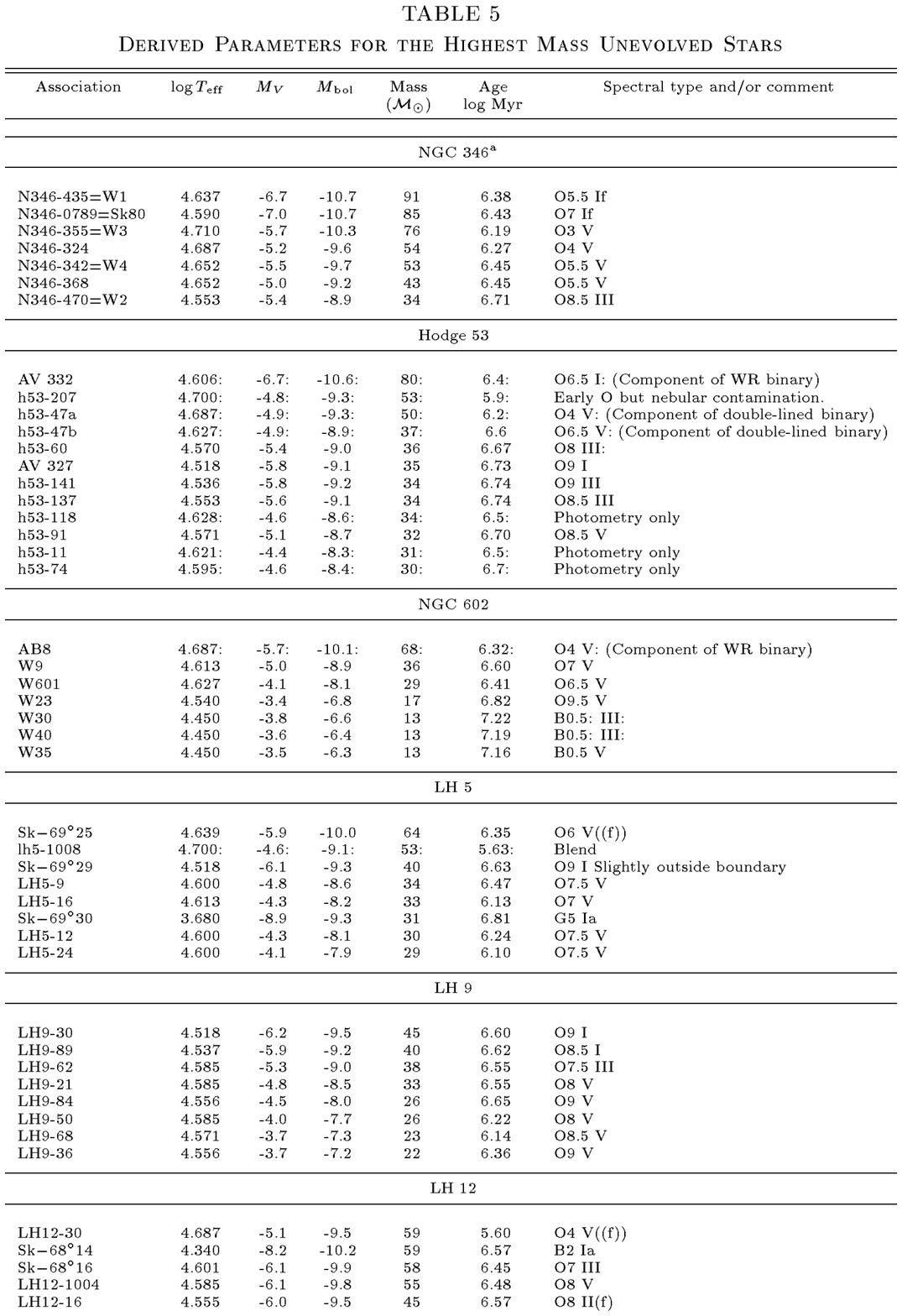}
\end{figure}

\clearpage
\begin{figure}
\epsscale{1.15}
\plotone{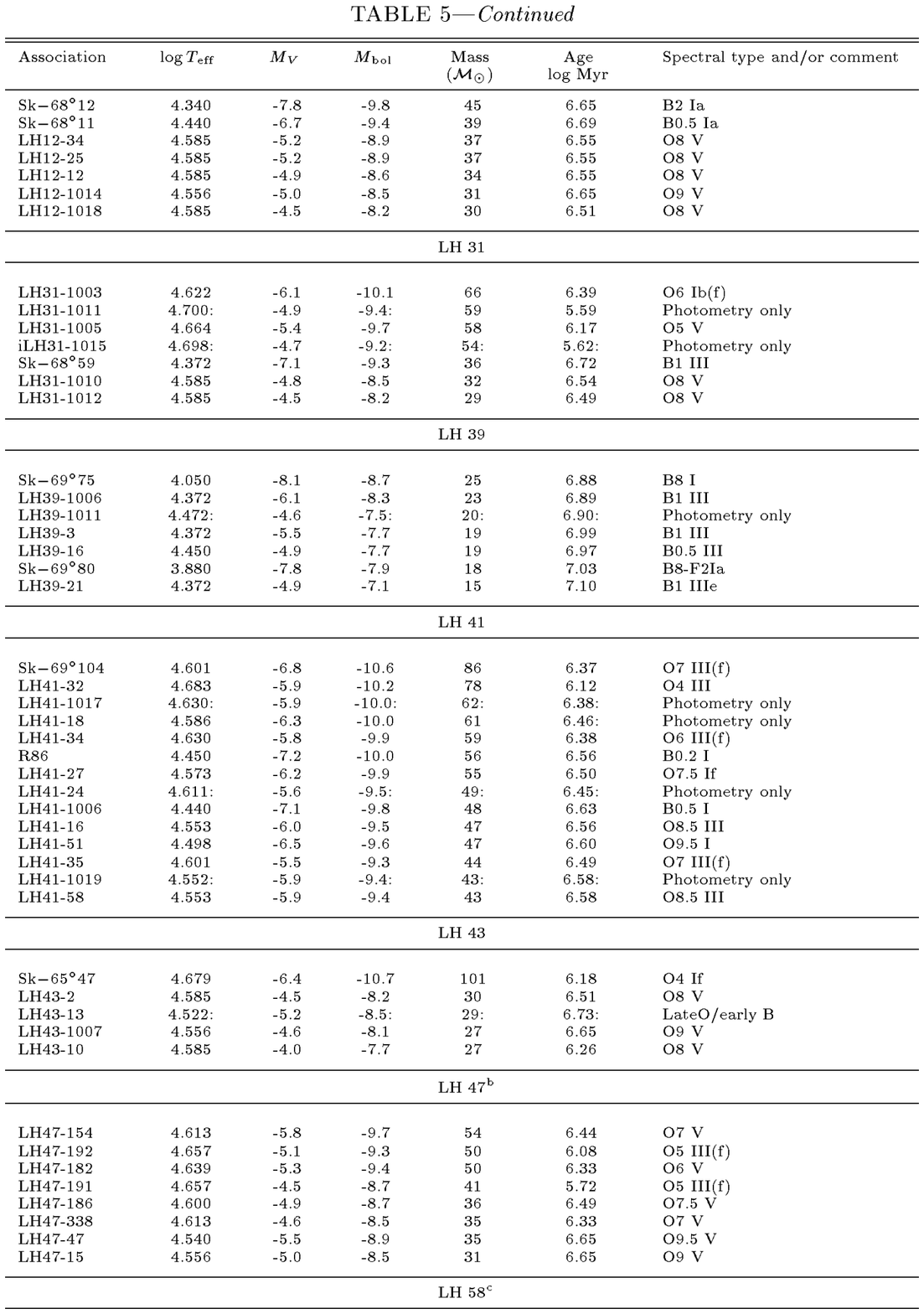}
\end{figure}

\clearpage
\begin{figure}
\epsscale{1.15}
\plotone{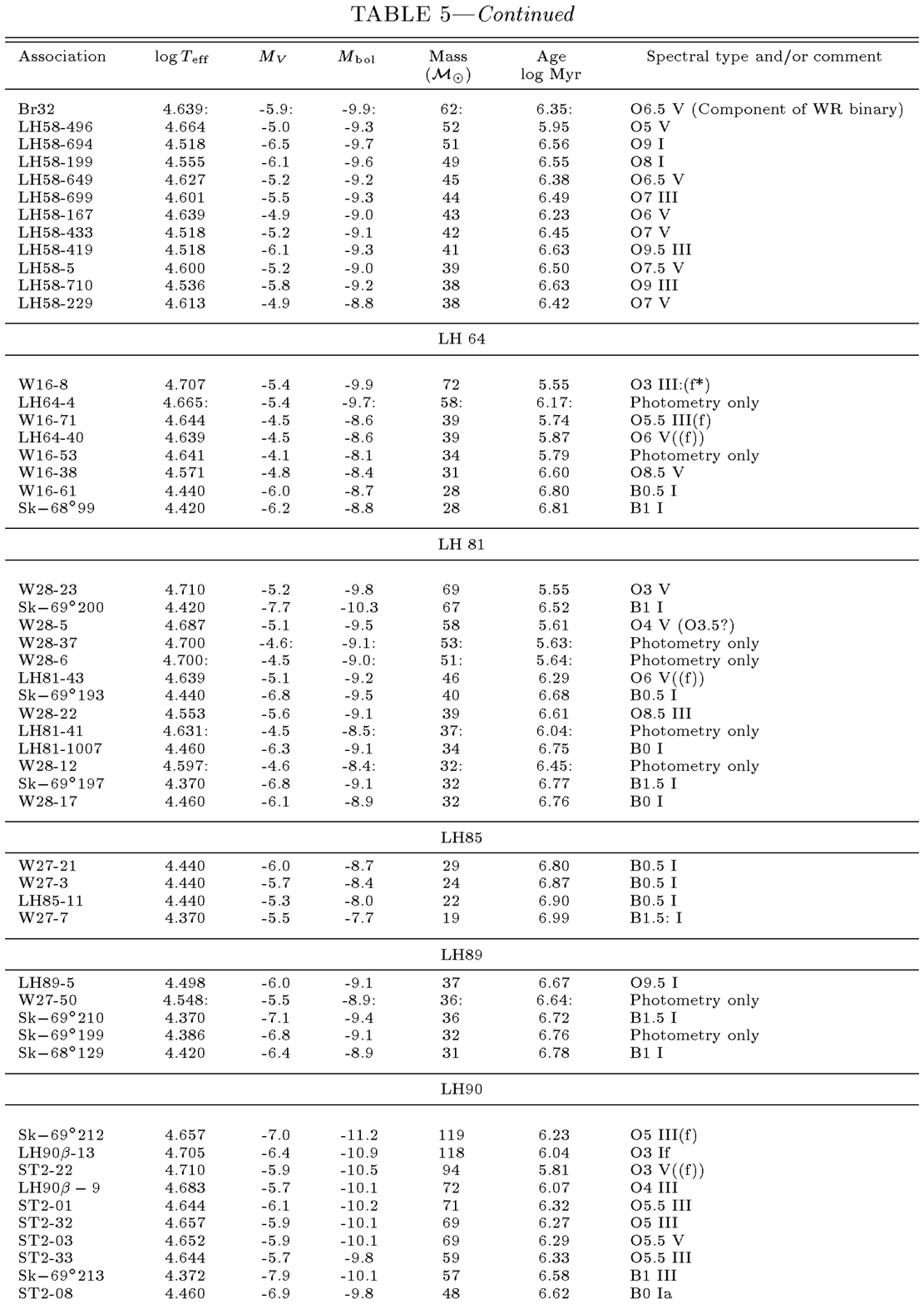}
\end{figure}

\clearpage
\begin{figure}
\epsscale{1.15}
\plotone{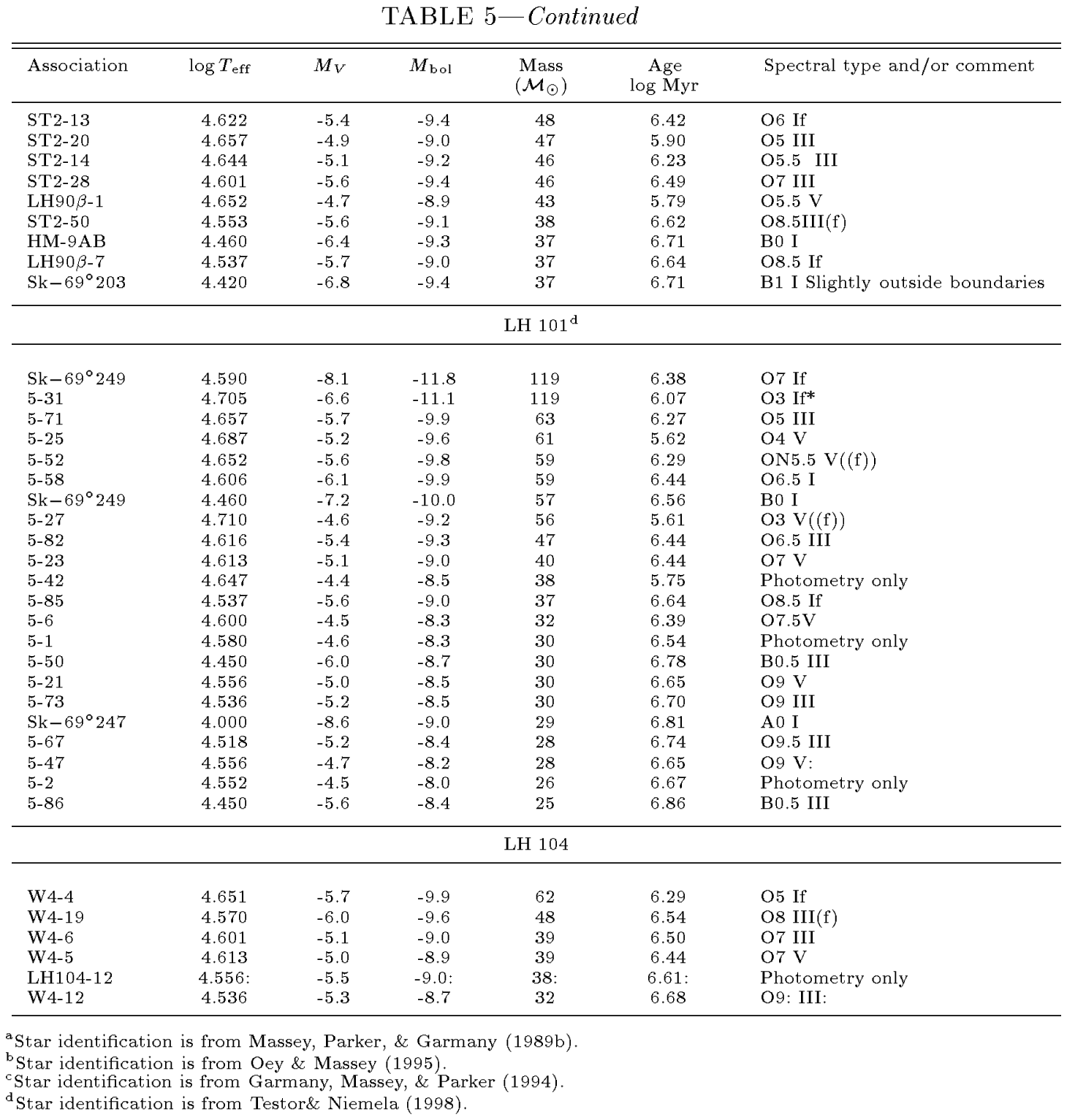}
\end{figure}

\clearpage
\begin{figure}
\epsscale{1.15}
\plotone{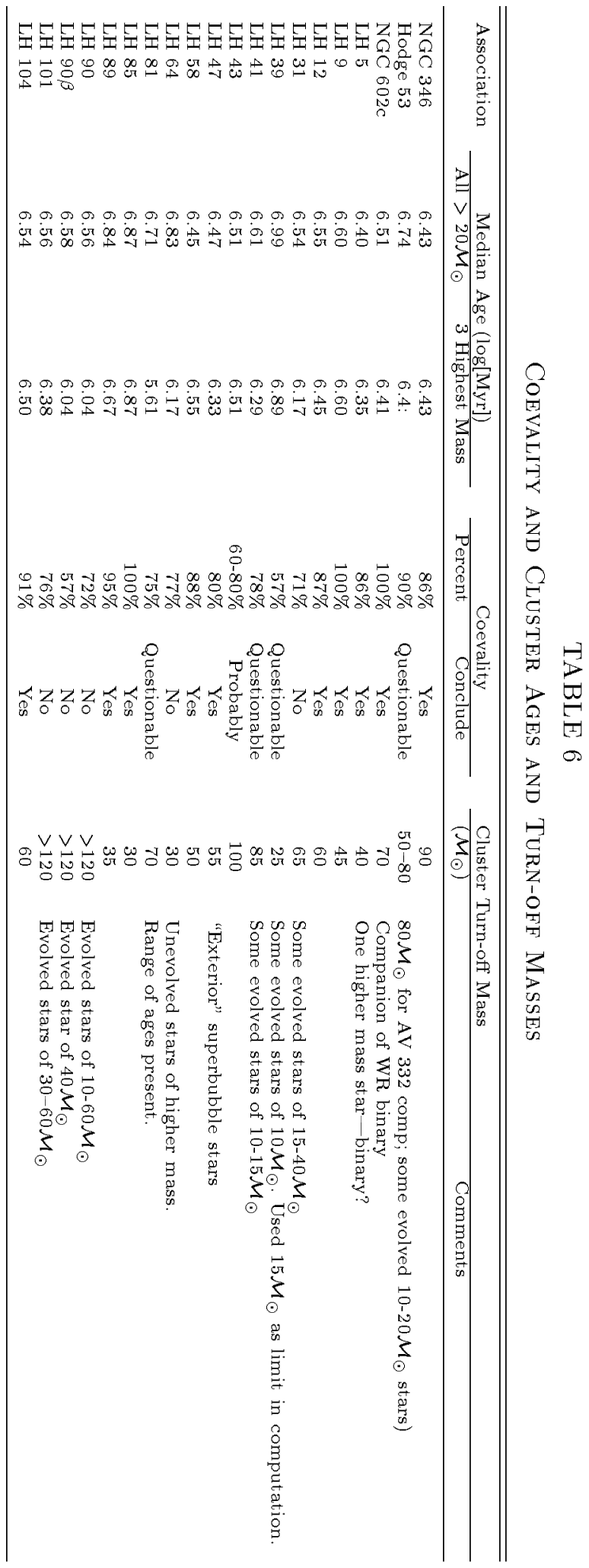}
\end{figure}

\clearpage
\begin{figure}
\epsscale{1.15}
\plotone{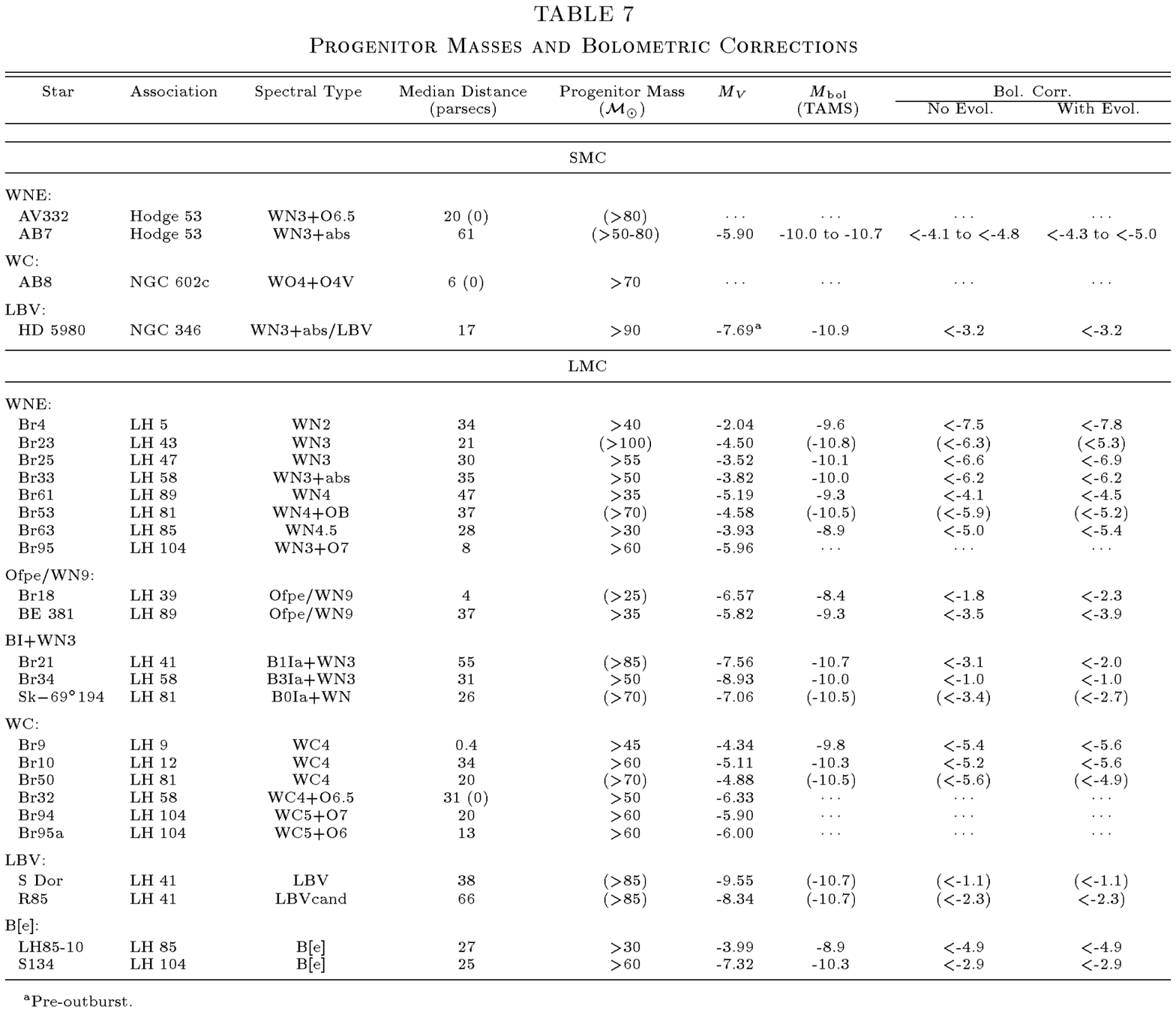}
\end{figure}

\clearpage

\figcaption{Three spectra of the suspected LBV R~85 are shown.  The star was classified by Feast et al.\ (1960) as an B5~Iae, roughly consistent with the 
spectrum we obtained in January 1999.  Spectra from two earlier times show a veiled appearance, with a spectral type that is cooler, based upon the lack of He~I $\lambda 4771$ compared to neighboring Mg~II $\lambda 4481$.
\label{fig:r85}}

\figcaption{The spectra of two O3~If* stars are shown
(LH90$\beta-13$ and ST5-31 in LH~101), along with
that of an O3~III(f*) star (W16-8 in LH~64).
\label{fig:o31}}

\figcaption{The spectra of two O3~V(f*) stars, ST2-22 in LH~90 [previously
classified as O3 ~III(f) by Testor et al.\ 1993],  and W28-23 
in LH~81.  The
third star, W28-5, also in LH~81, appears to be intermediate between O3~V and
O4~V, as the He~I $\lambda 4471$ strength would imply an O4 
classification, while the presence of N~V $\lambda 4603,19$ 
absorption would suggest an O3 description.
\label{fig:o32}}

\figcaption{The spectra of several early O-type dwarfs are show.
\label{fig:oms}}

\figcaption{The spectra of several O-type supergiants are shown.
\label{fig:osgs}}

\figcaption{The star Br~58 in LH~90 
has previously been called a WR star of type
WN5-6 or WN6-7.  We suggest here that it may be better described as one of
the H-rich transition objects of type O3~If*/WN6, i.e., an O3~If* star that
is so luminous that its stellar wind has come to resemble a WR star. (See
discussion in Massey \& Hunter 1998.)  The B0I+WN star W28-10, in LH~81, is
newly discovered here.
\label{fig:wrs}}

\figcaption{The H-R diagrams for the 19 OB associations studied here are shown. Stars for which spectral types were available are shown by
filled circles; stars for which only photometry was available are shown by
open circles. Asterisks represent stars with spectral types but whose location
in the HRD is considered particularly uncertain, usually the components of
spectroscopic binaries.  The location of the stars
denoted by the ``+" symbol are particularly
uncertain in the HRD. The solid lines show the evolutionary tracks for
the various (initial) masses as indicated.  The dashed lines are isochrones
at 2~Myr, 4~Myr, 6~Myr, and 10~Myr.  The tracks and isochrones come from
the $z=0.001$ models of Schaller et al.\ (1992) for the SMC associations, and
for the $z=0.008$ models of Schaerer et al.\ (1993) for the LMC associations.
\label{fig:hrds}}

\figcaption{How much of an error in age or mass is made by misclassifying a
star by a single spectral type?  The tracks and isochrones shown in these
HRDs are the same as in Fig.~\ref{fig:hrds} computed for LMC metallicity.
In (a) we show explicitly the discontinuities and gaps associated with 
adjacent spectral classification, as well as the systematic deviation
from the ZAMS at lower masses.  The upper sequence (supergiants) include
spectral 
types O3, O4, O5, O5.5, O6, O6.5, O7, O7.5, O8, O8.5, O9, O9.5, B0, B0.2,
B0.5, B1, B1.5, B2, B3, B5, B8, A0, A2, A5, A9, and F2.
The middle sequence (giants) include the same spectral types, but
terminating at B2.  The bottom sequence (dwarfs)
include the same sequence as the supergiants, but terminating at B3.
In (b) we show the errors that would
result for a misclassification by a single spectral subtype and/or
luminosity class for representative
points drawn from (a).  The points shown correspond to 
O3~I, O6~I, O8~I, B0~I, B1.5~I, B8~I, and A5~I among the upper sequence. 
The four giants shown in the middle sequence are: O5.5~III, O7.5~III, O9.5~III,
and B1~III. The five dwarfs shown along the bottom sequences are:
O4~V, O6.5~V, O8.5~V, B0.2~V, and B2~V. The error bars extend considerably
further than adjacent points in (a) because we have also included the
possibility of misclassification by a luminosity class; e.g., the possibility
that a star classified as an O7~III might actually be an O8~V.
\label{fig:err}}

\end{document}